\documentclass{article}

\title{Quantizing Ho\v{r}ava-Lifshitz Gravity\\ via Causal Dynamical Triangulations}

\author{\\ Christian Anderson${}^a$\footnote{canderson@college.harvard.edu} ,
Steven J. Carlip${}^b$\footnote{carlip@physics.ucdavis.edu} ,
Joshua H. Cooperman${}^b$\footnote{cooperman@physics.ucdavis.edu} ,
\smallskip\\
Petr Ho\v{r}ava${}^{c,d}$\footnote{horava@berkeley.edu} , Rajesh K. Kommu${}^b$\footnote{kommu@physics.ucdavis.edu} , and Patrick R. Zulkowski${}^{c,d}$\footnote{pzulkowski@berkeley.edu}
\\ \\
\emph{${}^a$Department of Physics, Harvard University, Cambridge, MA 02138}
\smallskip\\
\emph{${}^b$Department of Physics, University of California, Davis, CA 95616}
\smallskip\\
\emph{${}^c$Berkeley Center for Theoretical Physics and Department of
Physics,}\\
\emph{University of California, Berkeley, CA 94720}
\smallskip\\
\emph{${}^d$Theoretical Physics Group, Lawrence Berkeley National Laboratory,
Berkeley, CA 94720} }

\date{ }

\usepackage{subfigure}
\usepackage{tabularx}
\usepackage{booktabs}
\usepackage{graphicx}
\usepackage{amsmath}
\usepackage{latexsym}
\usepackage{mathrsfs}
\usepackage{geometry}
\numberwithin{equation}{section}

\begin{document}

\maketitle

\begin{abstract}

We extend the discrete Regge action of causal dynamical triangulations to include discrete versions of the curvature squared terms appearing in the continuum action of $(2+1)$-dimensional projectable Ho\v{r}ava-Lifshitz gravity. Focusing on an ensemble of spacetimes whose spacelike hypersurfaces are $2$-spheres, we employ Markov chain Monte Carlo simulations to study the path integral defined by this extended discrete action. We demonstrate the existence of known and novel macroscopic phases of spacetime geometry, and we present preliminary evidence for the consistency of these phases with solutions to the equations of motion of classical Ho\v{r}ava-Lifshitz gravity.  Apparently, the phase diagram contains a phase transition between a time-dependent de Sitter-like phase and a time-independent phase.  We speculate that this phase transition may be understood in terms of deconfinement of the global gravitational Hamiltonian integrated over a spatial $2$-sphere.

\end{abstract}
\vfill\break

\section{Connections}

An intriguing body of evidence hinting at a deep connection between Ho\v{r}ava-Lifshitz gravity%
\footnote{We have decided, by a majority vote among the coauthors of this paper, to follow the terminology commonly accepted in the literature and refer to gravity models with anisotropic scaling as Ho\v{r}ava-Lifshitz gravity, despite the dissenting vote of one of us (PH).}
and causal dynamical triangulations has recently accumulated in the literature. First, one of us demonstrated consistency of the spectral dimension computed in Ho\v{r}ava-Lifshitz gravity with the spectral dimension measured in causal dynamical triangulations in $3+1$ dimensions \cite{PetrSpec}. Benedetti \emph{et al} then verified within causal dynamical triangulations \cite{SpecDimHB} the prediction of Ho\v{r}ava-Lifshitz gravity for the behavior of the spectral dimension in $2+1$ dimensions \cite{PetrSpec}. Next, Ambj\o rn \emph{et al} exhibited the remarkable resemblance of the phase diagram of causal dynamical triangulations \cite{CDTandHL} to the phase diagram of Lifshitz matter systems \cite{PetrYM} as exemplified by the Lifshitz scalar field \cite{GenCovar}. These authors and one of us also conjectured \cite{CDTandHL,GenCovar} that the apparent tricritical point of the former phase diagram could correspond to the tricritical limit of Ho\v{r}ava-Lifshitz gravity with dynamical critical exponent $z$ equal to the dimension of space. Then, both Benedetti \emph{et al} and Ambj\o rn \emph{et al} noted the compatibility of certain solutions of Ho\v{r}ava-Lifshitz gravity with the minisuperspace model fit to the expectation value of the geometries emerging from causal dynamical triangulations \cite{CDTandHL,SpecDimHB}. Recently, Sotiriou \emph{et al} successfully fit the behavior of the spectral dimension of causal dynamical triangulations at intermediate scales to a dispersion relation derived from Ho\v{r}ava-Lifshitz gravity \cite{Visser3}. Just days ago, Budd argued that the kinetic term in the semiclassical effective action of causal dynamical triangulations exhibits a Ho\v{r}ava-Lifshitz-like form \cite{Budd}.

This mounting evidence motivated us to extend the Regge action---or, more precisely, the discrete path integral measure---used in causal dynamical triangulations to include dependence on the broader class of terms appearing in the action of Ho\v{r}ava-Lifshitz gravity. Our purpose is threefold: firstly, to test the applicability of causal dynamical triangulations to modified classical theories of gravitation; secondly, to explore quantum Ho\v{r}ava-Lifshitz gravity with nonperturbative techniques; and, thirdly, to illuminate further the links between Ho\v{r}ava-Lifshitz gravity and causal dynamical triangulations. As an initial step toward these goals, we have begun to investigate an appropriate reduction of $(2+1)$-dimensional projectable Ho\v{r}ava-Lifshitz gravity using causal dynamical triangulations. This model, though removed from the phenomenologically interesting case of $3+1$ dimensions, serves as a simplified yet nontrivial beginning for our research.

After briefly introducing the formalisms of Ho\v{r}ava-Lifshitz gravity and causal dynamical triangulations in section \ref{background}, we explain our adaptation of causal dynamical triangulations to Ho\v{r}ava-Lifshitz gravity in section \ref{discreteaction}. We present the results of our initial numerical studies---evidence for the existence of extended phases of geometry, the structure of these phases, and their relation to the classical solutions---in section \ref{results}. Finally, we summarize our conclusions and discuss ongoing and future work in section \ref{outlook}. To streamline our presentation, we relegate to appendix \ref{app1} the construction of the relevant classical solutions of Ho\v{r}ava-Lifshitz gravity and to appendix \ref{app2} certain geometric properties of causal dynamical triangulations.

\section{Background}\label{background}

\subsection{Ho\v{r}ava-Lifshitz Gravity}\label{Horava}

Ho\v{r}ava-Lifshitz gravity is a field theory of the dynamical metric on spacetime manifolds that carry the additional structure of a preferred foliation by spacelike hypersurfaces. The relevant set of gauge symmetries is then the group $\mbox{Diff}_{F}(\mathcal{M})$ of diffeomorphisms of the spacetime manifold $\mathcal{M}$ that preserve the preferred foliation $F$.  The foliation structure leads to one important novelty: the possibility of anisotropic scaling with a nontrivial dynamical scaling exponent $z$ measuring the degree of anisotropy between space and time. In Minkowski spacetime the relativistic scaling is thus replaced by the anisotropic scaling
\begin{subequations}
\begin{eqnarray}
 t&\longrightarrow& \tilde{t}=b^zt\\
\mathbf{x}&\longrightarrow& \tilde{\mathbf{x}}=b\mathbf{x}
\end{eqnarray}
\end{subequations}
for constant $b>0$. Starting with $z>1$ at short distances markedly improves the ultraviolet behavior of the theory, potentially rendering it power-counting renormalizable.

One might be tempted to rewrite this theory in a manifestly relativistic fashion by integrating in the remaining part of the group $\mbox{Diff}(\mathcal{M})$ of full spacetime diffeomorphisms. This procedure leads to a reformulation of the model as a specific scalar-tensor theory with higher-derivative interactions in which the spacelike hypersurfaces of constant scalar field dynamically determine the leaves of the foliation $F$.  Such a relativistic rewriting is only equivalent to the original nonrelativistic formulation at the classical level with, moreover, subtle regularity conditions on the scalar field's dynamics. At the quantum level the relativistic rewriting immediately besets the theory with the notorious problem of time, whereas the nonrelativistic nature of the original formulation of Ho\v{r}ava-Lifshitz gravity potentially renders the spacetime metric's dynamics more directly compatible with quantum mechanics. Additionally, as we shall show, the original formulation of Ho\v{r}ava-Lifshitz gravity is nicely suited to the framework of causal dynamical triangulations, which also utilizes a preferred foliation structure in the microscopic definition of the path integral for gravity.

In a local smooth coordinate chart $(t,\mathbf{x})$ adapted to the preferred foliation $F$, $\mbox{Diff}_{F}(\mathcal{M})$ consists of all reparametrizations of the form
\begin{subequations}\label{reparametrizations}
\begin{eqnarray}
t&\longrightarrow&\tilde{t}=f(t)\label{timereparametrizations}\\
\mathbf{x}&\longrightarrow&\tilde{\mathbf{x}}=\zeta(t,\mathbf{x})
\end{eqnarray}
\end{subequations}
for arbitrary functions $f$ and $\zeta$. Note that, although reparametrizations of the space coordinates may be time-dependent, reparametrizations of the time coordinate must be space-independent. Given its preferred foliation, the structure of Ho\v{r}ava-Lifshitz gravity is naturally discussed in the Arnowitt-Deser-Misner formalism \cite{ADM}. In this formalism the spacetime metric tensor $\mathbf{g}(t,\mathbf{x})$ is decomposed into the metric tensor $\mathbf{\gamma}(t,\mathbf{x})$ on a spacelike hypersurface $\Sigma$ of constant time coordinate $t$, the shift vector $\mathbf{N}(t,\mathbf{x})$, and the lapse function $N(t,\mathbf{x})$ such that one can reassemble the standard line element as
\begin{equation}\label{relatds}
\mathrm{d}s^{2}=-N^{2}(t,\mathbf{x})\mathrm{d}t^{2}+\gamma_{ij}(t,\mathbf{x})\left[\mathrm{d}x^{i}+N^{i}(t,\mathbf{x})\mathrm{d}t\right]\left[\mathrm{d}x^{j}+N^{j}(t,\mathbf{x})\mathrm{d}t\right].
\end{equation}
Note, however, that recombining the spatial metric tensor, the shift vector, and the lapse function into the spacetime line element (\ref{relatds}) is somewhat misleading since, in the regime with $z>1$, different terms contributing to $\mathrm{d}s^2$ carry different scaling dimensions.

The shift vector and the lapse function play the role of gauge fields associated with $\mbox{Diff}_{F}(\mathcal{M})$. Since the time coordinate reparametrizations \eqref{timereparametrizations} are independent of the space coordinates, a natural choice is to restrict the corresponding gauge field $N(t,\mathbf{x})$ also to be only a function of the time coordinate. We choose to make this restriction, yielding the so-called projectable version of the theory. (See, for instance, \cite{LifGrav,GBHealthy,Sot,Visser,Wang}.) Of course, we have interest in studying the more general nonprojectable version in which the lapse function is a spacetime field and of which the projectable version is a dynamical limit. The gauge symmetries then permit new terms in the action involving spatial derivatives of the lapse function; finding the appropriate realization of such terms in the framework of causal dynamical triangulations is beyond the present work's scope. We instead focus on the projectable version of Ho\v{r}ava-Lifshitz gravity, which has a particularly clear translation into the language of causal dynamical triangulations.

Employing the Arnowitt-Deser-Misner decomposition of the metric tensor, we now construct the action of projectable Ho\v{r}ava-Lifshitz gravity. We aim to build a power-counting renormalizable unitary classical theory of gravitation. Even in relativistic gravity, including higher curvature terms can render the theory renormalizable \cite{Hamber,Stelle}; typically, however, such terms come at the cost of sacrificing perturbative unitarity and propagating unphysical degrees of freedom \cite{Salam}. These issues stem from the inclusion of higher temporal derivative terms, which necessarily accompany higher spatial derivative terms in a relativistic theory. By only permitting higher spatial derivatives, we can in principle avoid problems with perturbative unitarity and unphysical degrees of freedom. Working in the Arnowitt-Deser-Misner formalism allows for the rather straightforward inclusion of higher spatial derivatives and exclusion of higher temporal derivatives as we desire in constructing the action of Ho\v{r}ava-Lifshitz gravity.

We build the general action in $d$ spatial dimensions as the sum of a kinetic term quadratic in temporal derivatives and a potential term of mass dimension $2d$ in spatial derivatives. The dynamical critical exponent $z$ thus equals $d$. Accordingly, we form the kinetic term from invariants of $\partial_{t}\mathbf{\gamma}(t,\mathbf{x})$. This derivative alone is not covariant under $\mbox{Diff}_{F}(\mathcal{M})$, but the extrinsic curvature tensor $\mathbf{K}(t,\mathbf{x})$ of a spacelike hypersurface $\Sigma$ satisfies this criterion. With $\mathbf{K}(t,\mathbf{x})$ having components
\begin{equation}
K_{ij}(t,\mathbf{x})=\frac{1}{2N(t)}\left[\partial_{t}{\gamma}_{ij}(t,\mathbf{x})-\nabla_{i}N_{j}(t,\mathbf{x})-\nabla_{j}N_{i}(t,\mathbf{x})\right]
\end{equation}
for the covariant derivative $\nabla$ associated with $\mathbf{\gamma}(t,\mathbf{x})$, the most general kinetic term invariant under $\mbox{Diff}_{F}(\mathcal{M})$ is
\begin{equation}\label{eq:kinpiece}
\frac{1}{16\pi G} \int_{\mathcal{M}}\mathrm{d}t\,\mathrm{d}^{d}x\sqrt{\gamma(t,\mathbf{x})}N(t)\left[K_{ij}(t,\mathbf{x})K^{ij}(t,\mathbf{x})-\lambda K^{2}(t,\mathbf{x})\right].
\end{equation}
Here the parameter $\lambda$ arises from the generalized DeWitt supermetric compatible with the theory's gauge symmetries, and $K(t,\mathbf{x})$ is the trace of the extrinsic curvature tensor $\mathbf{K}(t,\mathbf{x})$ \cite{QuantMem,LifGrav}.

We form the potential term from invariants of $\mathbf{\gamma}(t,\mathbf{x})$ and its spatial derivatives. The most general potential term is
\begin{equation}\label{eq:potpiece}
\frac{1}{16\pi G}\int_{\mathcal{M}}\mathrm{d}t\,\mathrm{d}^{d}x\sqrt{\gamma(t,\mathbf{x})}N(t)V\left[\mathbf{\gamma}(t,\mathbf{x})\right],
\end{equation}
where $V\left[\mathbf{\gamma}(t,\mathbf{x})\right]$ is a scalar functional of the spatial metric tensor and its spatial derivatives up to order $2d$. Putting together \eqref{eq:kinpiece} and \eqref{eq:potpiece}, the action of projectable Ho\v{r}ava-Lifshitz gravity is
\begin{equation}
S_{HL}[\mathbf{g}(t,\mathbf{x})]=\frac{1}{16\pi G} \int_{\mathcal{M}}\mathrm{d}t\,\mathrm{d}^{d}x\sqrt{\gamma(t,\mathbf{x})}N(t)\left\{K_{ij}(t,\mathbf{x})K^{ij}(t,\mathbf{x})-\lambda K^{2}(t,\mathbf{x}) - V\left[\mathbf{\gamma}(t,\mathbf{x})\right]\right\}.
\end{equation}
Note that the coupling constant $G$ is related but not equal to the Newton constant $G_N$: the low energy Newton constant $G_N$ is determined by rescaling the time coordinate by an effective speed of light factor selected so that the term linear in the spatial Ricci scalar within $V\left[\mathbf{\gamma}(t,\mathbf{x})\right]$ is correctly normalized with respect to the kinetic term \cite{QuantMem,LifGrav}.  This rescaling also dictates how the constant term in $V\left[\mathbf{\gamma}(t,\mathbf{x})\right]$ relates to the low energy cosmological constant $\Lambda$.

For $d>2$, with the anticipated $z=d$ scaling at short distances, there is a proliferation of marginal and relevant contributions to the potential term; for $d=2$, however, the most general action containing only the marginal and relevant terms for $z=2$ is quite compact:
\begin{eqnarray}\label{HLaction}
S_{HL}[\mathbf{g}(t,\mathbf{x})]&=&\frac{1}{16\pi G}\int_{\mathcal{M}}\mathrm{d}t\,\mathrm{d}^{2}x\sqrt{\gamma(t,\mathbf{x})}N(t)\left[K_{ij}(t,\mathbf{x})K^{ij}(t,\mathbf{x})-\lambda K^{2}(t,\mathbf{x})\right.\nonumber\\ && \left.-\alpha R_{2}^{2}(t,\mathbf{x})+\beta R_{2}(t,\mathbf{x})-2\Lambda \right]
\end{eqnarray}
with $R_{2}(t,\mathbf{x})$ the Ricci scalar of the spatial metric tensor $\mathbf{\gamma}(t,\mathbf{x})$ and the coupling constant $\Lambda$ related to the low energy cosmological constant $\Lambda$ by the rescaling described above. The potential term's relative simplicity in $(2+1)$-dimensional projectable Ho\v{r}ava-Lifshitz gravity further motivates our initially studying this case.

The equations of motion \eqref{eq:metriceom} derived from variation of the action \eqref{HLaction}  with respect to the spatial metric tensor are independent of the coupling constant $\beta$: for $d=2$ the $R_{2}$ term is a total derivative, and the Gauss-Bonnet theorem determines its integral over a spacelike hypersurface $\Sigma$ of the preferred foliation in terms of the Euler number of $\Sigma$.  Similarly, the equations of motion \eqref{eq:momconstraint} obtained from variation of the action \eqref{HLaction}  with respect to the shift vector are also insensitive to $\beta$. The coupling constant $\beta$ does, however, appear in the equation of motion derived from variation of the action \eqref{HLaction} with respect to the lapse function. Since the lapse function only depends on the time coordinate, its equation of motion takes the form of a spatially integrated Hamiltonian constraint
\begin{equation}\label{eq:intconstraint}
{\mathcal{H}}_\perp=0
\end{equation}
for the global Hamiltonian
\begin{equation}\label{hperp}
{\mathcal{H}}_\perp=\int_{\Sigma}\mathrm{d}^{2}x \sqrt{\gamma(t,\mathbf{x})} \left[K_{ij}(t,\mathbf{x})K^{ij}(t,\mathbf{x})-\lambda K^{2}(t,\mathbf{x}) + \alpha R_{2}^{2}(t,\mathbf{x}) - \beta R_{2}(t,\mathbf{x}) + 2 \Lambda \right].
\end{equation}

At this stage there is an additional choice to make beyond that of restricting to projectable Ho\v{r}ava-Lifshitz gravity. Recall the distinction that arises between noncompact and compact spatial topology when imposing the local Hamiltonian constraint in relativistic theories of gravity. In the former case, after imposing appropriate asymptotic fall-off conditions and accounting for possible boundary contributions, the zero mode ${\mathcal{H}}_\perp$ of the local Hamiltonian constraint does not vanish on physical states. Instead, this mode is one of the conserved asymptotic charges, namely the total energy. In the latter case the zero mode ${\mathcal{H}}_\perp$ of the local Hamiltonian constraint vanishes on all physical states since the relativistic gauge symmetries mix it together with the other gauge symmetry generators.

Now, in Ho\v{r}ava-Lifshitz gravity, as pointed out in \cite{GenCovarC,GenCovar,CenkePetr}, this situation is somewhat different: while the case of noncompact spatial topology remains essentially unchanged, a novelty arises in the case of compact spatial topology. Imposing the vanishing of $\mathcal{H}_\perp$ on physical states is not mandatory: either one may treat $\mathcal{H}_\perp$ as a gauge symmetry generator and thus impose the condition ${\mathcal{H}}_\perp=0$, or one may treat ${\mathcal{H}}_\perp$ as a conserved global charge measuring the energy levels of physical states in the Hilbert space. Of course, the second option requires us to check that this Hamiltonian's spectrum is bounded from below on physical states. This novel situation was explicitly encountered in \cite{CenkePetr} whose authors constructed from bosonic systems on a rigid lattice the renormalization group fixed points corresponding to the free field limit of Ho\v{r}ava-Lifshitz gravity with $z=2$ and $z=3$. There the specific choices that one could make at the level of the microscopic lattice Hamiltonian gave rise to the option of imposing or not imposing the condition $\mathcal{H}_\perp=0$ for the effective low energy gravitons.

When we discretize the action \eqref{HLaction} in section \ref{discreteaction} for use as a functional on simplicial manifolds, we find that the second option---treating ${\mathcal{H}}_\perp$ as a conserved global charge---appears more natural in the setting of causal dynamical triangulations. The geometric restrictions derived by employing simplices of fixed edge lengths suggest that the lapse function has been effectively set to a constant, though without leading in any discernible way to the imposition of the constraint \eqref{eq:intconstraint}. Also, the $R_{2}$ term with coupling constant $\beta$ in the action \eqref{HLaction} does not affect any observables for the topological reasons stated above. The value of $\mathcal{H}_\perp$ does, however, depend on $\beta$, which suggests that enforcing the constraint \eqref{eq:intconstraint} is not completely consistent. In section \ref{physsemi}, where we compare the geometries emerging from our numerical simulations to the classical solutions for the constraint \eqref{eq:intconstraint} not imposed, we find evidence supporting our selection of the second option. These comparisons are, moreover, not without weight: as briefly discussed in \cite{GenCovar} and further demonstrated in \cite{Zulkowski}, the classical phase diagram for the theory with variable lapse differs significantly from that of the present setting. The classical theory that we quantize in section \ref{results} is thus defined by the action \eqref{HLaction} for fixed lapse.

\subsection{Causal Dynamical Triangulations}\label{CDT}

The causal dynamical triangulations approach aspires to define the continuum limit of a quantum theory of gravity by appealing solely to those nonperturbative tools applied with great success to the quantization of local gauge field theories. (See \cite{kogut1,kogut2} for reviews.) In particular, the approach invokes a lattice regularization of the path integral for gravity and then utilizes the renormalization group to explore its continuum limit. In causal dynamical triangulations one thus attempts to define this path integral as the partition function of a statistical model of dynamical geometry. (See \cite{Entropic,LatUpdate} for recent reviews.) As in several previous programs of lattice quantization of gravity, the statistical ensemble of geometries is comprised of simplicial manifolds triangulated by a fixed set of $(d+1)$-simplices. In $2+1$ dimensions this set consists of those $3$-simplices or tetrahedra pictured in figure $1$.
\begin{figure}
\centering
\label{fig:tetra}
\includegraphics[scale=0.5]{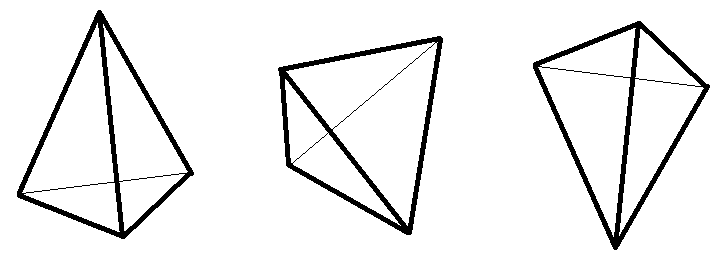}
\caption{From left to right: the $(3,1)$-, $(2,2)$-, and $(1,3)$-tetrahedra. The abscissa indicates the number of vertices on an initial triangulated spacelike hypersurface, and the ordinate indicates the number of vertices on the next triangulated spacelike hypersurface.}
\end{figure}

The novel feature of causal dynamical triangulations lies in the imposition of an additional restriction on the geometries permitted to enter the path integral: these simplicial manifolds must possess a global foliation by spacelike hypersurfaces of constant discrete time. The global foliation was originally introduced to enable a Wick rotation from Lorentzian to Riemannian signature. The causal structure of a Lorentzian triangulation could thus be faithfully retained in its Wick rotated version, only the latter triangulation being amenable to numerical analysis \cite{AJL1,DynTri2001,AJL2}. A simplicial manifold is endowed with the foliation structure as follows: every spacelike hypersurface, all of a chosen fixed topology, is triangulated with equilateral $d$-simplices, and then the vertices of adjacent spacelike hypersurfaces are connected by timelike edges so as to produce only the $(d+1)$-simplices of the fixed set. Spacelike edges have length squared $l_{SL}^{2}=a^{2}$, defining a lattice spacing for the triangulation, and timelike edges have length squared $l_{TL}^{2}=-\eta a^{2}$ for $\eta>0$ \cite{AJL1,DynTri2001}. Note that our parameter $\eta$ corresponds to the parameter $\alpha$ typically used in the causal dynamical triangulations literature.

More concretely, one aims to approximate the path integral for general relativity,
\begin{equation}\label{grpath}
Z\left[\mathbf{\gamma}(t_{f},\mathbf{x})|\mathbf{\gamma}(t_{i},\mathbf{x})\right]=\int_{\mathbf{\gamma}(t_{i},\mathbf{x})}^{\mathbf{\gamma}(t_{f},\mathbf{x})}\mathcal{D}\mathbf{g}(t,\mathbf{x})\,e^{iS_{EH}[\mathbf{g}(t,\mathbf{x})]},
\end{equation}
by the path sum
\begin{equation} \label{eq:partfnc}
Z\left[\Gamma_{f}|\Gamma_{i}\right]=\sum_{\mathcal{T}}\frac{1}{C_{\mathcal{T}}}\,e^{iS_{R}[\mathcal{T}]}.
\end{equation}
In \eqref{grpath} the path integral is taken over all physically distinct spacetime metric tensors $\mathbf{g}(t,\mathbf{x})$ interpolating between the initial and final boundary geometries specified by the spatial metric tensors $\mathbf{\gamma}(t_{i},\mathbf{x})$ and $\mathbf{\gamma}(t_{f},\mathbf{x})$, and $S_{EH}$ denotes the Einstein-Hilbert action
\begin{equation}\label{EHaction}
S_{EH}[\mathbf{g}(t,\mathbf{x})]=\frac{1}{16\pi G_N}\int_{\mathcal{M}}\mathrm{d}t\,\mathrm{d}^{d}x\,\sqrt{-g(t,\mathbf{x})}\left[R(t,\mathbf{x})-2\Lambda\right].
\end{equation}
In \eqref{eq:partfnc} the path sum is taken over all causal triangulations $\mathcal{T}$ interpolating between the initial and final boundary geometries specified by the triangulations $\Gamma_{i}$ and $\Gamma_{f}$ \cite{AJL1,DynTri2001}, and $S_{R}$ denotes the Regge action
\begin{equation}\label{Reggeaction}
S_{R}[\mathcal{T}]=\frac{1}{8\pi G_N}\sum_{h\in\mathcal{T}}A_{h}\delta_{h}-\frac{\Lambda}{16\pi G_N}\sum_{s\in\mathcal{T}}V_{s}.
\end{equation}
Here, $h$ is a $(d+1-2)$-dimensional hinge having area $A_{h}$ and deficit angle $\delta_{h}$, and $V_{s}$ is the spacetime volume of a $(d+1)$-simplex \cite{Regge}. The measure factor $\frac{1}{C_{\mathcal{T}}}$ is the inverse of the order of the automorphism group of $\mathcal{T}$, included to account for discrete symmetries arising in the structure of $\mathcal{T}$.

Before proceeding further with our discussion of causal dynamical triangulations, we make a few clarifying remarks on the implications of the above prescription for regularizing the continuum path integral \eqref{grpath}. The restriction to simplicial manifolds admitting a preferred foliation effectively changes the path integral measure and integration domain. Even though we continue to employ the Regge action, these changes may result in our model belonging to a universality class different from that of Euclidean dynamical triangulations.  In $1+1$ dimensions, where the path sum \eqref{eq:partfnc} can be evaluated analytically, this is known to be the case \cite{ACJL}: by preventing the birth of baby universes, the foliation requirement alters the critical exponents, indicating that Euclidean and causal dynamical triangulations occupy different universality classes. In higher dimensions the same situation seems very likely to hold true; otherwise, we should expect our model to fall into one of the universality classes in which no smooth macroscopic limit exists, and the spacetime geometry exhibits a branched polymer or crumpled behavior.

The change in the path integral measure and integration domain may also translate into a change in our theory's effective action such that it no longer assumes the general relativistic form. Recall that, although we may locally treat the choice of a preferred foliation as a gauge choice, we cannot generally make such a choice globally. A gauge choice typically introduces a Jacobian into the path integral measure, and a mismatch between the path integral measure and the gauge choice can lead to a breaking of the gauge symmetry. Starting with the Regge action, the question of whether counterterms sensitive to the preferred foliation are generated thus remains open. If such counterterms appear, then the resulting effective action likely corresponds to Ho\v{r}ava-Lifshitz gravity at some particular values of its couplings. The Regge action's bare coupling constants $G_{N}$ and $\Lambda$ are only indirectly related to the continuum renormalized values of the Newton constant and the cosmological constant at long distances, so we must not naively identify the former with the latter.

Continuing with our discussion of causal dynamical triangulations, the action \eqref{Reggeaction} simplifies considerably for the simplicial manifolds contributing to our ensemble since only a small set of simplices with fixed geometries is used to construct them. Still, the path sum \eqref{eq:partfnc} has resisted all attempts at analytical computation for $d>1$, so exploration of its properties has primarily employed numerical techniques. In particular, Markov chain Monte Carlo methods are used to simulate the path sum \eqref{eq:partfnc} \cite{Semiclassical,2ndorder,3Dnonpert,AJL2,AJL3,AJL4,AJL5,AJL6,JK2,JK1}. To render the path sum \eqref{eq:partfnc} amenable to such analysis, we must first Wick rotate the real time action \eqref{Reggeaction} to imaginary time, a well-defined process owing to the foliated structure of causal dynamical triangulations. We perform the Wick rotation by analytically continuing the parameter $\eta$ in the lower half complex plane \cite{DynTri2001}. For our case of  interest---$2+1$ dimensions with spacelike hypersurfaces having the topology of $\mathcal{S}^{2}$ and periodic time coordinate having the topology of the $1$-sphere $\mathcal{S}^{1}$---the Regge action becomes
\begin{equation}\label{CDTaction}
S_{CDT}^{(E)}=-k_{0}N_{0}+k_{3}N_{3}
\end{equation}
after Wick rotation and application of various topological relations for $\eta=1$ \cite{3Dnonpert}. Here, $N_{0}$ is the number of vertices and $N_{3}$ is the number $3$-simplices in the triangulation $\mathcal{T}$. The coupling constants $k_{0}$ and $k_{3}$ are related to the bare couplings $G_N$ and $\Lambda$ as
\begin{subequations}
\begin{eqnarray}
k_{0}&=&\frac{a}{4G_N} \\
k_{3}&=&\frac{a^{3}\Lambda}{48\sqrt{2}\pi G_N}+\frac{a}{4G_N}\left(\frac{3}{\pi}\cos^{-1}{\frac{1}{3}}-1\right),
\end{eqnarray}
\end{subequations}
We have thus transformed the path sum \eqref{eq:partfnc} into the statistical partition function
\begin{equation}\label{ECDTpartfnc}
Z\left[\Gamma_{f}|\Gamma_{i}\right]=\sum_{\mathcal{T}}\frac{1}{C_{\mathcal{T}}}\,e^{-S_{CDT}^{(E)}[\mathcal{T}]}.
\end{equation}

Computer simulations of the partition function \eqref{ECDTpartfnc} in both $2+1$ and $3+1$ dimensions have thus far provided considerable support for the existence of an extended phase of geometry possessing not only a semiclassical limit on large scales, but also a quantum regime on small scales \cite{Semiclassical,3Dnonpert,AJL2,AJL3,AJL4,SpecDimAmbjorn,AJL5,AJL6,SpecDimHB,JK2,JK1}. In particular, the average observed geometry matches well that of (possibly deformed) Euclidean de Sitter spacetime at both the classical and semiclassical levels \cite{Semiclassical,3Dnonpert,AJL2,AJL3,AJL4,SpecDimAmbjorn,AJL5,AJL6,SpecDimHB,JK2}. Furthermore, studies of the spectral dimension of this extended phase of geometry have revealed an apparent dimensional reduction to effective $2$-dimensionality on small scales \cite{SpecDimAmbjorn,SpecDimHB,JK1}. This phenomena of dynamical dimensional reduction---particularly, the extrapolated value of the minimal dimensionality---has elicited comparisons of causal dynamical triangulations to both the asymptotic safety approach and Ho\v{r}ava-Lifshitz gravity. In the former theory the effective change in the spectral dimension apparently results from large anomalous dimensions near the nontrivial fixed point, while in the latter theory the spectral dimension flows to the short distance value of 2 as a result of the anisotropic scaling near the Gaussian fixed point \cite{PetrSpec,FracSpace}. Further evidence for such dimensional reduction has also surfaced in other approaches to the quantization of general relativity \cite{Carlip}.

As in Euclidean dynamical triangulations, additional phases of geometry also emerge \cite{EDT1}. At first these phases were viewed as unphysical, but current interpretations favor their role as further phases in the vicinity of a multicritical fixed point \cite{CDTandHL}. Specifically, in both $ 2+1 $ and $3+1$ dimensions there exists a phase characterized by spacelike hypersurfaces that effectively decouple from one another \cite{3Dnonpert,AJL4,JK1}, and in $3+1$ dimensions a second additional phase, distinguished by its large Hausdorff dimension, appears \cite{AJL4,JK1}. Recently, Ambj\o rn \emph{et al} have argued that the transition between this last phase and the physical phase is of second order \cite{2ndorder}. This finding raises the possibility of rigorously defining a continuum limit of causal dynamical triangulations.

\section{A Discrete Action for Ho\v{r}ava-Lifshitz Gravity}\label{discreteaction}

We now derive a discrete form of the action \eqref{HLaction} suitable for analysis with the techniques of causal dynamical triangulations. As above we assume a topology of $\mathcal{S}^{2}\times\mathcal{S}^{1}$, primarily motivated by the relative ease of numerically analyzing such compact spacetimes.  As in the lattice regularizations of local quantum field theories \cite{kogut1,kogut2}, the precise details of the discretization do not matter since universality ensures that many of the details at the lattice spacing scale become irrelevant in the long distance limit. Our primary goal in constructing a discrete analogue of the action \eqref{HLaction} is thus to build an action sufficiently specific so that the continuum limit lies in the universality class of Ho\v{r}ava-Lifshitz gravity yet sufficiently generic so that the continuum limit does not depend on all of the renormalized coupling constants in this universality class. Of course, we can only judge whether or not we have achieved these goals once we have thoroughly studied the quantum theory of the model defined below.

Working from the continuum action \eqref{HLaction}, we allow two technical criteria to guide our construction of its discrete version: first, the discrete action should manifestly reduce to the Regge action used in causal dynamical triangulations when the bare coupling constants $\lambda$ and $\alpha$ assume their general relativistic values, and, second, the transfer matrix corresponding to the discrete action defined on the space of boundary geometries should yield a well-defined Hamiltonian. These two criteria apply to the classical discrete action constructed below. In the quantum theory defined via the path integral based on this action, we naturally anticipate a nontrivial relationship between the bare coupling constants and the renormalized coupling constants. Depending on this relationship, the quantum theory's long distance continuum limit may or may not be relativistic.

To implement the first criterion, we use the Gauss-Codazzi equation,
\begin{equation}
R(t,\mathbf{x})=R_{2}(t,\mathbf{x})-\left[K^{2}(t,\mathbf{x})-K_{ij}(t,\mathbf{x})K^{ij}(t,\mathbf{x})\right]+\mathrm{Total}\,\mathrm{Derivative},
\end{equation}
to rewrite the action \eqref{HLaction} as
\begin{eqnarray}\label{eq:HLrewrite}
S_{HL}[\mathbf{g}(t,\mathbf{x})]&=&\frac{1}{16\pi G}\int_{\mathcal{M}}\mathrm{d}t\,\mathrm{d}^{2}x\,\sqrt{-g(t,\mathbf{x})}\left[R(t,\mathbf{x})-2\Lambda\right]\nonumber\\ && + \frac{1-\lambda}{16\pi G} \int_{\mathcal{M}}\mathrm{d}t\,\mathrm{d}^{2}x \sqrt{\gamma(t,\mathbf{x})} N(t)K^{2}(t,\mathbf{x})\nonumber\\ && -\frac{\alpha}{16\pi G} \int_{\mathcal{M}} \mathrm{d}t\,\mathrm{d}^{2}x \sqrt{\gamma(t,\mathbf{x})} N(t) R_{2}^{2}(t,\mathbf{x})
\end{eqnarray}
up to irrelevant boundary terms. We neglect the term in the action \eqref{HLaction} with coupling constant $\beta$ as it contributes only an additive constant for the spacetime manifolds under consideration. In this form the action \eqref{eq:HLrewrite} straightforwardly reduces to the Einstein-Hilbert action \eqref{EHaction} when the coupling constants $\lambda$ and $\alpha$ take on their  general relativistic values of one and zero. We use the discrete action \eqref{CDTaction} for the Einstein-Hilbert portion of the action \eqref{eq:HLrewrite}, leaving us the task of determining the discrete analogues of the $K^{2}$ and $R_{2}^{2}$ terms. We construct these terms below, postponing discussion of the second criterion.

\subsection{Curvature Squared Terms in Regge Calculus}\label{curvature2}

There exist well-established prescriptions for constructing the Ricci scalar and the trace of the extrinsic curvature tensor in Regge calculus. The former is defined in terms of deficit angles $\delta_{h}$ about $(d+1-2)$-dimensional hinges $h$ of a $(d+1)$-dimensional simplicial manifold \cite{Regge}:
\begin{equation}\label{R}
\int_{\mathcal{M}}\mathrm{d}t\,\mathrm{d}^{d}x\,\sqrt{-g(t,\mathbf{x})}R(t,\mathbf{x})=2\sum_{h\in\mathcal{T}}A_{h}\delta_{h}.
\end{equation}
The latter is defined in terms of angles $\psi_{h}$ between the normal vectors to the two $d$-simplices intersecting at the $(d+1-2)$-dimensional hinge $h$ within the spacelike hypersurface $\Sigma$ of a $(d+1)$-dimensional simplicial manifold \cite{Brewin,KinRegge}:
\begin{equation}\label{K}
\int_{\Sigma}\mathrm{d}^{d}x\,\sqrt{\gamma(t,\mathbf{x})}K(t,\mathbf{x})=\sum_{h\in\Sigma}A_{h}\psi_{h}.
\end{equation}
Technically, these curvatures have a distributional definition on the hinges, which complicates the construction of curvature squared terms. In particular, simply taking the square of \eqref{R} or \eqref{K} to define the discrete versions of the $R_{2}^{2}$ or $K^{2}$ terms leads to a mathematically ill-defined continuum limit \cite{AmbjornHD}. Accordingly, we adhere to the philosophy of \cite{AmbjornHD,HamberWilliams} when building discrete analogues of curvature squared terms, adopting the alternative scheme of volume sharing. As discussed in \cite{AmbjornHD}, we  make the identification
\begin{equation}\label{volsharecurvature}
\int_{\Sigma}\mathrm{d}^{d}x\sqrt{\gamma(t,\mathbf{x})}\mathscr{R}^{2}(t,\mathbf{x})\longrightarrow \sum_{o \in O_{\tau}(\mathcal{T})} V_{o}^{(s)} \left( \frac{\delta_{o} V_{o}}{V_{o}^{(s)}} \right)^{2}
\end{equation}
for a curvature scalar $\mathscr{R}(t,\mathbf{x})$, where $o$ is the object assigned the curvature density, $O_{\tau}(\mathcal{T})$ is the set of all objects $o$ on the spacelike hypersurface $\Sigma$ labelled by discrete time coordinate $\tau$ in the triangulation $\mathcal{T}$, $V_{o}$ is the appropriate volume of the object $o$, and $V_{o}^{(s)}$ is the share-volume of the object $o$, namely the volume of all top-dimensional objects containing $o$. Using this scheme we now address the $R_{2}^{2}$ and $K^{2}$ terms in turn.

\subsubsection{$R_{2}^{2}$ Term}\label{R2}

For the $ R_{2}^{2} $ term there are clear choices for the objects $o$ and the top-dimensional objects containing $o$: since the Ricci scalar characterizes the intrinsic geometry of a $2$-dimensional spacelike hypersurface, the objects $o$ are vertices $v$ and the top-dimensional objects are spacelike triangles $\triangle$. This is completely consistent with the usual prescription for the Ricci scalar in Regge calculus. For a vertex $v$ the deficit angle is
\begin{equation}
\delta_{v}=2 \pi - \frac{\pi}{3} N_{\triangle}(v)
\end{equation}
for the number $N_{\triangle}(v)$ of spacelike triangles containing $v$, the volume is
\begin{equation}
V_{v}=1,
\end{equation}
and the share-volume is
\begin{equation}
V_{v}^{(s)} = \sum_{\triangle\supset v} A_{\triangle} = \frac{\sqrt{3}}{4} a^2 N_{\triangle}(v)
\end{equation}
since all of the spacelike triangles are equilateral. Employing \eqref{volsharecurvature} and noting that the most natural discretization of the time integral is
\begin{equation}
\int\mathrm{d}t\,N(t) \longrightarrow \sum_{\tau } \sqrt{\eta} a,
\end{equation}
we make the identification
\begin{equation}\label{eqn:discR2}
\int_{\mathcal{M}}\mathrm{d}t\,\mathrm{d}^{2}x \sqrt{\gamma(t,\mathbf{x})} N(t) R_{2}^{2}(t,\mathbf{x}) \longrightarrow\sum_{\tau}   \sum_{v\in V_{\tau}(\mathcal{T})} \frac{\sqrt{\eta}}{a}\frac{\left(6-N_{\triangle}(v)\right)^{2}}{N_{\triangle}(v)}.
\end{equation}
up to multiplicative factors, where $ V_{\tau}(\mathcal{T}) $ denotes the set of vertices $v$ belonging to the spacelike hypersurface labeled by the discrete time coordinate $ \tau $.

\subsubsection{$K^{2}$ Term}\label{K2}

For the $K^{2}$ term the choices of objects $o$ and top-dimensional objects containing $o$ are not as clear. The extrinsic curvature captures in part how the spacelike hypersurface is embedded in the spacetime manifold. The discrete analogue of the $ K^{2} $ term must reflect this geometric information, requiring that we appropriately account for how tetrahedra connect to the spacelike hypersurface. This observation suggests that we take tetrahedra as the top-dimensional objects contributing to the share-volume. Furthermore, in the continuum $ K^{2} $ scales as an inverse length squared, implying that the objects $ o $ be spacelike triangles.
\begin{figure}
\centering
\includegraphics[scale=0.5]{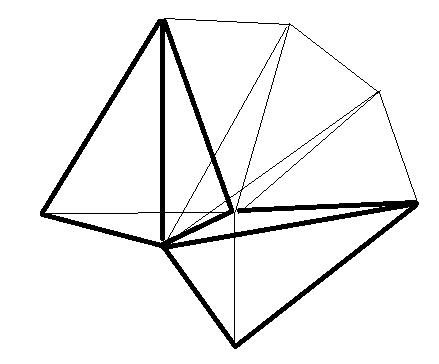}
\caption{An embedding in three dimensions of two $(3,1)$-tetrahedra (solid black) joined by three $(2,2)$- tetrahedra (thin black) all sharing a common edge. A vector perpendicular to the triangular base of a $(3,1)$-tetrahedron rotates through an angle $ \pi- 2\theta_{L}^{(3,1)}- 3 \theta_{L}^{(2,2)} $ as it is parallel transported across the common edge.}
\label{fan}
\end{figure}
We thus need to assign a deficit angle $ \delta_{\triangle} $ to a spacelike triangle. We largely follow the treatment of  \cite{Miller}. Consider a spacelike hypersurface of the triangulation  $\mathcal{T}$. For each spacelike triangle we may define a future-directed normal vector at the center of the $(3,1)$-tetrahedron of which it forms the base. A natural measure of extrinsic curvature at the common edge $e$ between two adjacent such spacelike triangles is the deficit of the angle that the normal vector traces out as it is parallel transported from its own $(3,1)$-tetrahedron to the adjacent $(3,1)$-tetrahedron. In figure $2$ we illustrate this construction for the particular case of three $(2,2)$-tetrahedra stacked between the two adjacent spacelike triangles. In general, the deficit angle for the common edge is
\begin{equation}
\delta_{e} = \frac{1}{i}\left(\pi- 2 \theta_{L}^{(3,1)} - \theta_{L}^{(2,2)} N^{\uparrow}_{(2,2)}(e)\right)
\end{equation}
with $ \theta_{L}^{(3,1)} $ and $ \theta_{L}^{(2,2)} $ the Lorentzian dihedral angles about spacelike edges and $ N^{\uparrow}_{(2,2)}(e) $ the number of future-directed $ (2,2) $-tetrahedra attached to the common edge $ e $. We give the values of  $ \theta_{L}^{(3,1)} $ and $ \theta_{L}^{(2,2)} $ in appendix \ref{app2}. We thus assign to a spacelike triangle the deficit angle
\begin{equation}
\delta_{\triangle} = \frac{1}{i}\left(3\pi- 6 \theta_{L}^{(3,1)} - \theta_{L}^{(2,2)} N^{\uparrow}_{(2,2)}(\triangle)\right),
\end{equation}
which, we note, is reminiscent of a trace. The volume of a spacelike triangle is
\begin{equation}
V_{\triangle}=\frac{\sqrt{3}}{4}a^{2},
\end{equation}
and the share volume is
\begin{equation}
V_{\triangle}^{(s)}=4V_{L}^{(3,1)}+V_{L}^{(2,2)}N_{(2,2)}^{\uparrow}(\triangle)
\end{equation}
with $V_{L}^{(3,1)}$ and $V_{L}^{(2,2)}$ the Lorentzian $3$-volumes of the respective tetrahedra.  We give the values of  $ V_{L}^{(3,1)} $ and $ V_{L}^{(2,2)} $ in appendix \ref{app2}. The share volume assumes this value since a given spacelike triangle has four $(3,1)$-tetrahedra and $N_{(2,2)}^{\uparrow}(\triangle)$ $(2,2)$-tetrahedra in its immediate future. According to the prescription \eqref{volsharecurvature}, we make the identification
\begin{equation}\label{discreteK2prelim}
\int_{\mathcal{M}}\mathrm{d}t\,\mathrm{d}^{2}x \sqrt{\gamma(t,\mathbf{x})} N(t) K^{2}(t,\mathbf{x}) \longrightarrow  \sum_{\tau}\sum_{\triangle \in T^{SL}_{\tau}(\mathcal{T})} a^{4}\frac{ \big| 3\pi- 6 \theta_{L}^{(3,1)} - \theta_{L}^{(2,2)} N^{\uparrow}_{(2,2)}(\triangle) \big|^{2} }{4 V_{L}^{(3,1)} + V_{L}^{(2,2)} N^{\uparrow}_{(2,2)}(\triangle)}
\end{equation}
up to multiplicative factors, where $T^{SL}_{\tau}(\mathcal{T})$ denotes the set of spacelike triangles $\triangle$ belonging to the spacelike hypersurface labelled by the discrete time coordinate $\tau$.

The discretization \eqref{discreteK2prelim} of the $K^{2}$ term does not, however, respect our second criterion. Following  \cite{DynTri2001}, to ensure the existence of a well-defined Hamiltonian on the space of boundary geometries, we must make \eqref{discreteK2prelim} time-reversal invariant. A straightforward calculation shows that this invariance guarantees positivity of the squared transfer matrix, which, along with the transfer matrix's symmetry, yields a well-defined Hamiltonian \cite{DynTri2001}. To realize time-reversal invariance, we add an analogous term for past-directed $(2,2)$-tetrahedra. The complete discrete analogue of the $ K^{2} $ term is
\begin{equation}\label{eq:discksqrd}
\sum_{\tau}\sum_{\triangle \in T^{SL}_{\tau}(\mathcal{T})}a^{4}\left[ \frac{\big| 3\pi- 6 \theta_{L}^{(3,1)} - \theta_{L}^{(2,2)} N^{\uparrow}_{(2,2)}(\triangle) \big|^{2}}{4 V_{L}^{(3,1)} + V_{L}^{(2,2)} N^{\uparrow}_{(2,2)}(\triangle)} +\frac{\big| 3\pi- 6 \theta_{L}^{(3,1)} - \theta_{L}^{(2,2)} N^{\downarrow}_{(2,2)}(\triangle) \big|^{2}}{4 V_{L}^{(3,1)} + V_{L}^{(2,2)} N^{\downarrow}_{(2,2)}(\triangle)} \right].
\end{equation}
Note that we did not require such an adjustment for the $R_{2}^{2}$ term since it depends only on the intrinsic geometry of the spacelike hypersurface.

\subsection{Imaginary Time Action}

Putting together  \eqref{eqn:discR2} and \eqref{eq:discksqrd}, our discrete action for Ho\v{r}ava-Lifshitz gravity becomes
\begin{eqnarray}\label{HLCDTaction}
S_{HL}[\mathcal{T}]&=&S_{CDT}[\mathcal{T}]\nonumber\\ &&+\frac{1-\lambda}{16\pi G} \sum_{\tau}\sum_{\triangle \in T^{SL}_{\tau}(\mathcal{T})}a^{4}\left[ \frac{\big| 3\pi- 6 \theta_{L}^{(3,1)} - \theta_{L}^{(2,2)} N^{\uparrow}_{(2,2)}(\triangle) \big|^{2}}{4 V_{L}^{(3,1)} + V_{L}^{(2,2)} N^{\uparrow}_{(2,2)}(\triangle)} +\frac{\big| 3\pi- 6 \theta_{L}^{(3,1)} - \theta_{L}^{(2,2)} N^{\downarrow}_{(2,2)}(\triangle) \big|^{2}}{4 V_{L}^{(3,1)} + V_{L}^{(2,2)} N^{\downarrow}_{(2,2)}(\triangle)} \right]\nonumber\\ &&-\frac{\alpha}{16\pi G} \sum_{\tau}\sum_{v\in V_{\tau}(\mathcal{T})}\frac{\sqrt{\eta}}{a} \frac{\left(6-N_{\triangle}(v)\right)^{2}}{N_{\triangle}(v)}.
\end{eqnarray}
Wick rotating the action \eqref{HLCDTaction} to imaginary time, we find that
\begin{eqnarray}\label{eq:discHLE}
S_{HL}^{(E)}[\mathcal{T}]&=&-k_{0}N_{0}+k_{3}N_{3}\nonumber\\ &&+\frac{1-\lambda}{16\pi G}  \sum_{\tau}\sum_{\triangle \in T^{SL}_{\tau}(\mathcal{T})}a^{4}\left[ \frac{\left( 3\pi- 6 \theta_{E}^{(3,1)} - \theta_{E}^{(2,2)} N^{\uparrow}_{(2,2)}(\triangle) \right)^{2}}{4 V_{E}^{(3,1)} + V_{E}^{(2,2)} N^{\uparrow}_{(2,2)}(\triangle)} +\frac{\left( 3\pi- 6 \theta_{E}^{(3,1)} - \theta_{E}^{(2,2)} N^{\downarrow}_{(2,2)}(\triangle) \right)^{2}}{4 V_{E}^{(3,1)} + V_{E}^{(2,2)} N^{\downarrow}_{(2,2)}(\triangle)} \right]\nonumber\\ && +\frac{\alpha}{16\pi G}  \sum_{\tau}\sum_{v\in V_{\tau}(\mathcal{T})}\frac{\sqrt{\eta}}{a} \frac{\left(6-N_{\triangle}(v)\right)^{2}}{N_{\triangle}(v)}.
\end{eqnarray}
We give the values of $ \theta_{E}^{(3,1)} $, $ \theta_{E}^{(2,2)} $, $ V_{E}^{(3,1)} $, and $ V_{E}^{(2,2)} $ in appendix \ref{app2}. We use the action \eqref{eq:discHLE} for $a=1$ and $\eta=1$ in the path integral quantization analyzed below. The value of the length $a$ has no \emph{a priori} physical meaning, and any value of the parameter $\eta>\frac{1}{2}$ is permitted.

\section{A Lattice Quantization of Ho\v{r}ava-Lifshitz Gravity}\label{results}

\subsection{Numerical Implementation}

Following the analyses of  \cite{Semiclassical,2ndorder,3Dnonpert,AJL2,AJL3,AJL4,AJL5,AJL6,JK2,JK1}, we explore the partition function
\begin{equation}\label{HLCDTpartfnc}
Z[\Gamma_{f}|\Gamma_{i}]=\sum_{\mathcal{T}}\frac{1}{C_{\mathcal{T}}}\,e^{-S_{HL}^{(E)}[\mathcal{T}]}
\end{equation}
employing Markov chain Monte Carlo methods. In our simulations so far, we have fixed the topology of the spacelike hypersurfaces to be that of $\mathcal{S}^{2}$ and have fixed the total number $T$ of spacelike hypersurfaces, introducing a discrete time coordinate $\tau$ that enumerates these hypersurfaces. In the simulations reported below, we have set $T=64$. Additionally, we impose periodic boundary conditions on this  time coordinate, endowing it with the topology of $\mathcal{S}^{1}$. Consequently, the initial triangulated spacelike hypersurface $\Gamma_{i}$ and the final triangulated spacelike hypersurface $\Gamma_{f}$ of each causal triangulations $\mathcal{T}$ entering the partition function \eqref{HLCDTpartfnc} are identified.

Next, we hold the number $N_{3}$ of tetrahedra in each triangulation $\mathcal{T}$ approximately fixed; otherwise, the weights $e^{-S_{HL}^{(E)}[\mathcal{T}]}$ appearing in the partition function \eqref{HLCDTpartfnc} may grow without bound, eventually causing our computer code to crash. To implement the constraint of fixed number of tetrahedra, we add to the action \eqref{eq:discHLE} a term of the form $\epsilon|N_{3}-\bar{N}_{3}|$, where $\epsilon$ is a Lagrange multiplier and $\bar{N}_{3}$ is the target number of tetrahedra. Essentially, this Lagrange multiplier term modifies the value of the coupling constant $k_{3}$. In all of the simulations reported below, we have set $\epsilon=0.02$ and $\bar{N}_{3}$ to approximately $10200$. These parameter values typically yield a one percent variation in $N_{3}$ over any given ensemble. This number of tetrahedra is sufficiently large for physical effects clearly to outweigh finite size effects and sufficiently small for our limited computing resources to survey a reasonable portion of the coupling constant space.

Finally, we tune to a set of coupling constants $\{k_{0},\lambda,\alpha,k_{3}^{c}(k_{0},\lambda,\alpha)\}$ on the critical surface of the coupling constant space defined by the partition function \eqref{HLCDTpartfnc} for a fixed $N_{3}$. As our notation suggests, we first select values for $k_{0}$, $\lambda$, and $\alpha$ and then tune to the associated critical value $k_{3}^{c}$ of $k_{3}$. The critical value $k_{3}^{c}$ is that for which $N_{3}$ remains approximately constant while a simulation runs. In this sense our model is only well-defined at the critical value: for any other value the number of tetrahedra either increases without bound or plummets to zero.

With these conditions established, a simulation begins with the generation of an initial triangulation having the topology $\mathcal{S}^{2}\times\mathcal{S}^{1}$ composed of $\bar{N}_{3}$ tetrahedra. Using the Pachner moves adapted to causal dynamical triangulations, as described, for instance, in \cite{DynTri2001}, we run a standard Metropolis algorithm to generate an ensemble of spacetimes representative of the weighting defined by the partition function \eqref{HLCDTpartfnc}, sampling only after a period of thermalization. We sample the representative spacetimes generated every one hundred sweeps, a single sweep comprising $\bar{N}_{3}$ attempted Pachner moves. Once collected, we estimate the expectation values of observables as averages over the ensemble.

Testing our code is a nontrivial matter: without a known nonperturbative quantization of $(2+1)$-dimensional projectable Ho\v{r}ava-Lifshitz gravity, we possess no standard of comparison for our results. Of course, this situation also pertains to causal dynamical triangulations formulated with the Regge action. Our code is a modified version of that reported in  \cite{JK1}, which has yielded independent corroboration of the results of  \cite{3Dnonpert,AJL2,AJL3,AJL4,SpecDimAmbjorn,AJL5,AJL6,SpecDimHB}. We have run the modified code at the general relativistic values of the coupling constants $\lambda$ and $\alpha$ to check that we correctly reproduce the results of these references. In figure $3$ we present depictions of two representative spacetimes---one in the physical phase and one in the decoupled phase of causal dynamical triangulations---generated by our code. In figure $4$ we plot the ensemble average spectral dimensions---discussed further below---for the two ensembles to which the representative spacetimes in figure $3$ belong. Up to finite size effects currently under investigation, these measurements agree quantitatively with those of  \cite{SpecDimHB,JK2,JK1}. In figure $8(a)$ below we also show the ensemble average discrete $2$-volume as a function of discrete time for the second of these two ensembles. This data, as well as the fit to it, are consistent with the findings of  \cite{3Dnonpert,JK2}. Beyond these checks, we rely on the plausibility of our new results as a test of our code's correctness.

\begin{figure}[ht]
\centering
\subfigure[ ]{
\includegraphics[scale=0.455]{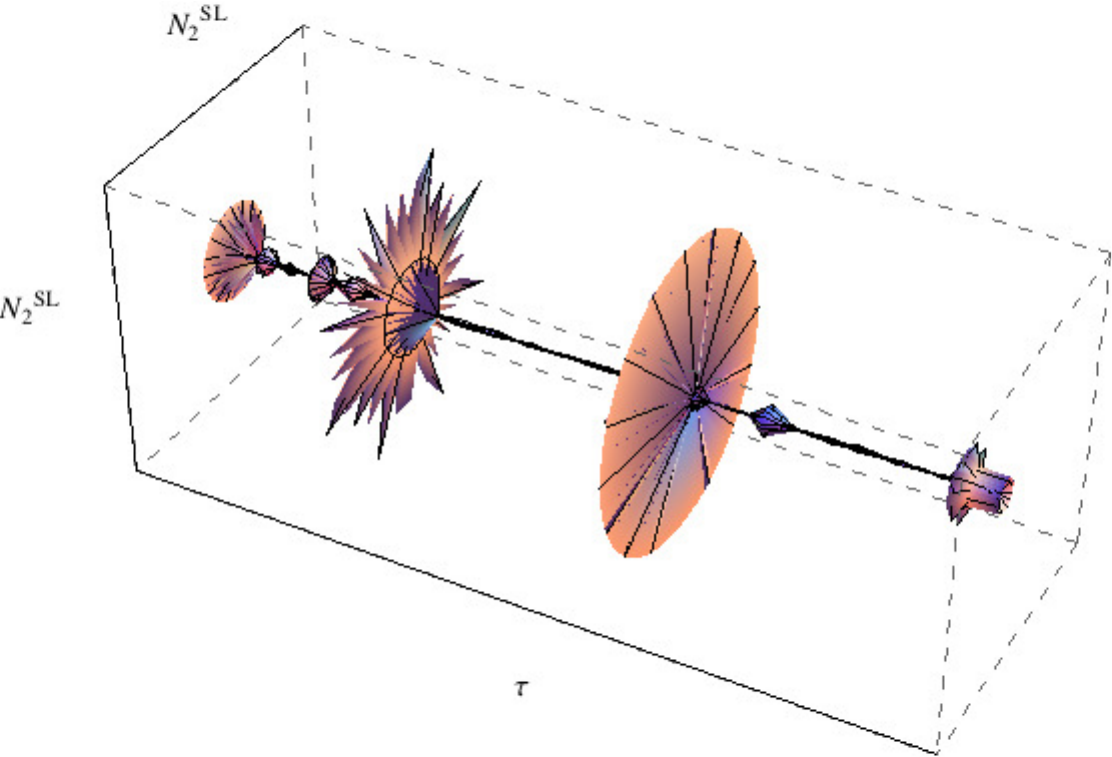}
\label{3d601000}
}
\subfigure[ ]{
\includegraphics[scale=0.455]{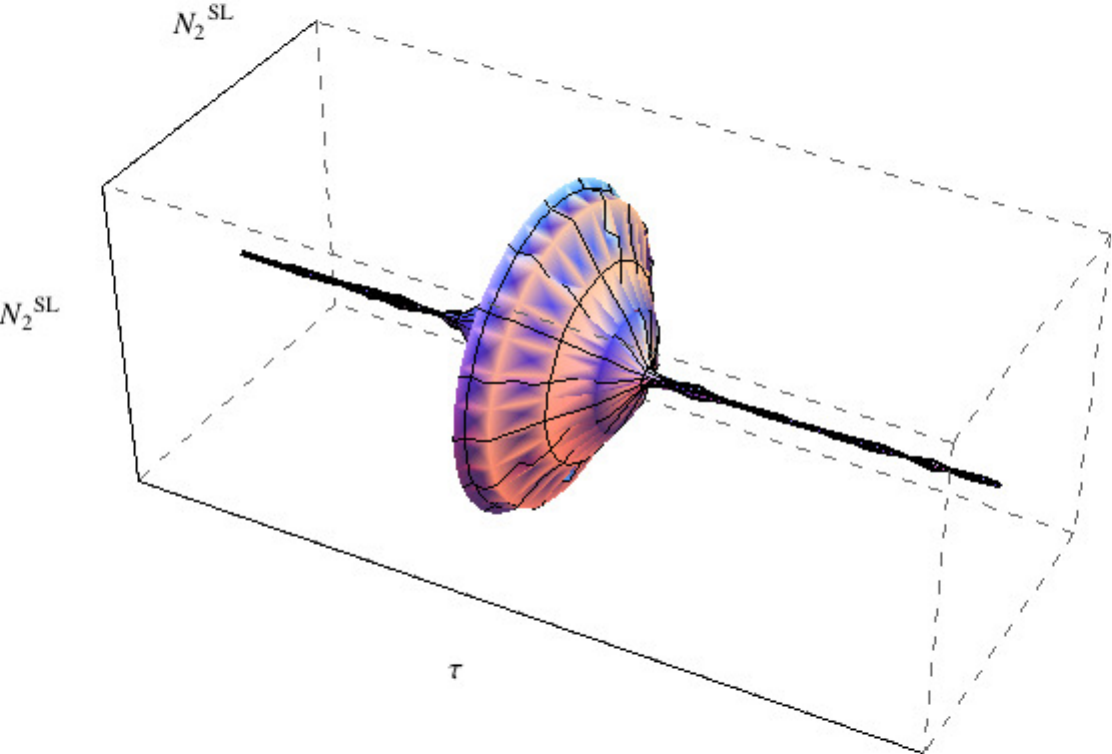}
\label{3d101000}
}
\\
\subfigure{
\includegraphics[scale=0.49]{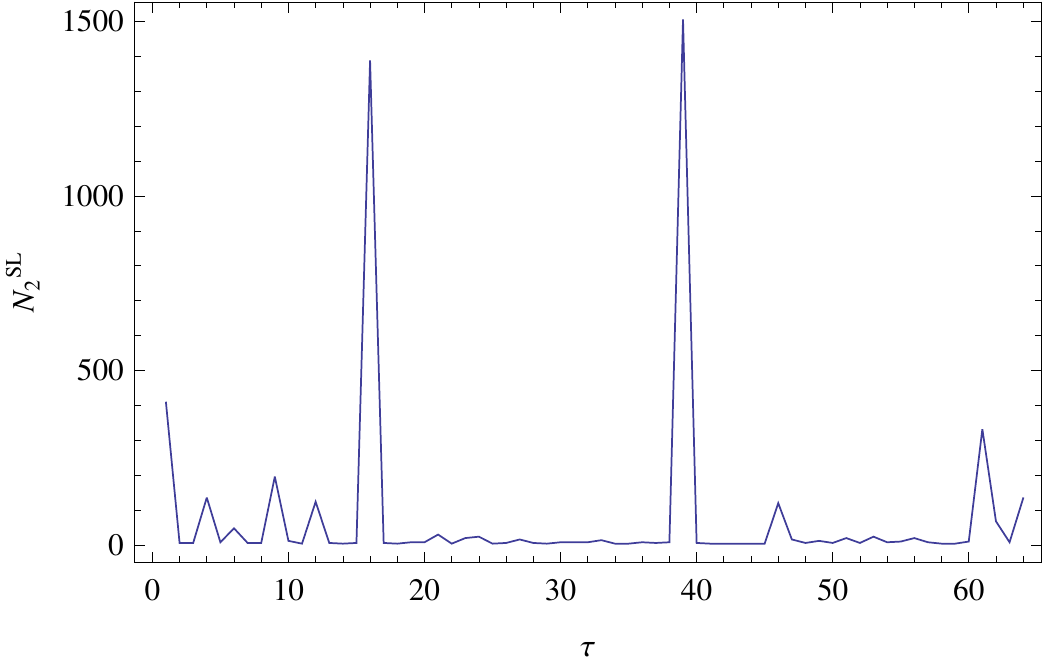}
\label{2d601000}
}
\subfigure{
\includegraphics[scale=0.49]{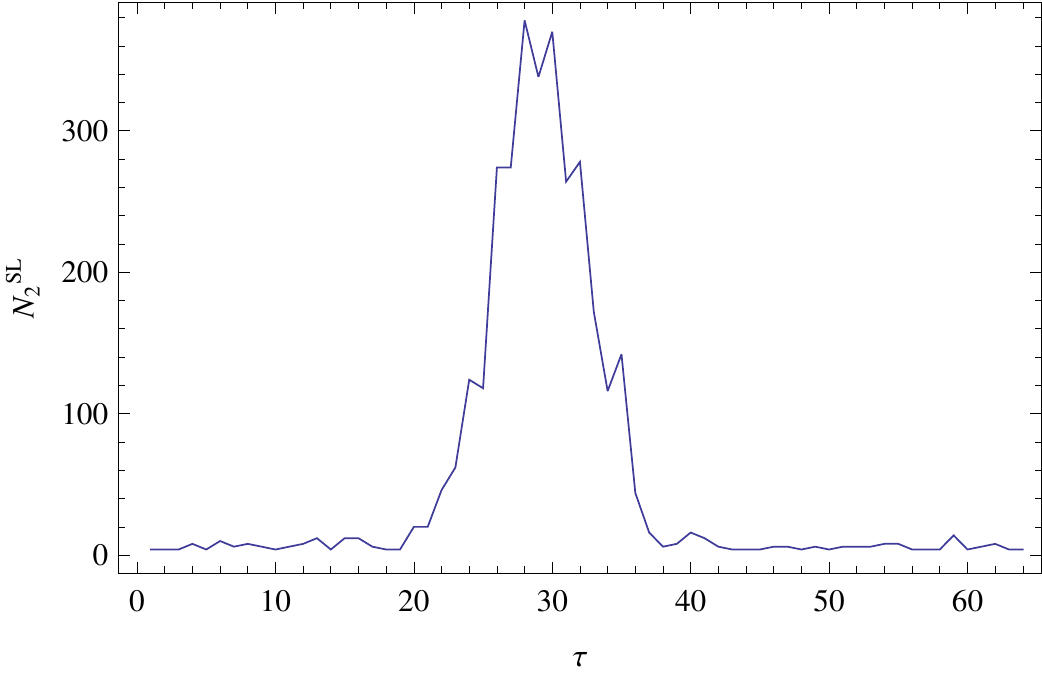}
\label{2d101000}
}
\label{CDTrepspacetimes}
\caption[Optional caption for list of figures]{Depictions of representative spacetimes showing the number $N_{2}^{SL}$ of spacelike triangles as a function of discrete time $\tau$.  \subref{3d601000} Phase A ($k_{0}=6.00$, $k_{3}=1.85$, $\lambda=1.00$, $\alpha=0.00$) \subref{3d101000} Phase C ($k_{0}=1.00$, $k_{3}=0.75$, $\lambda=1.00$, $\alpha=0.00$)}
\end{figure}

\begin{figure}[ht]
\centering
\subfigure[ ]{
\includegraphics[scale=0.49]{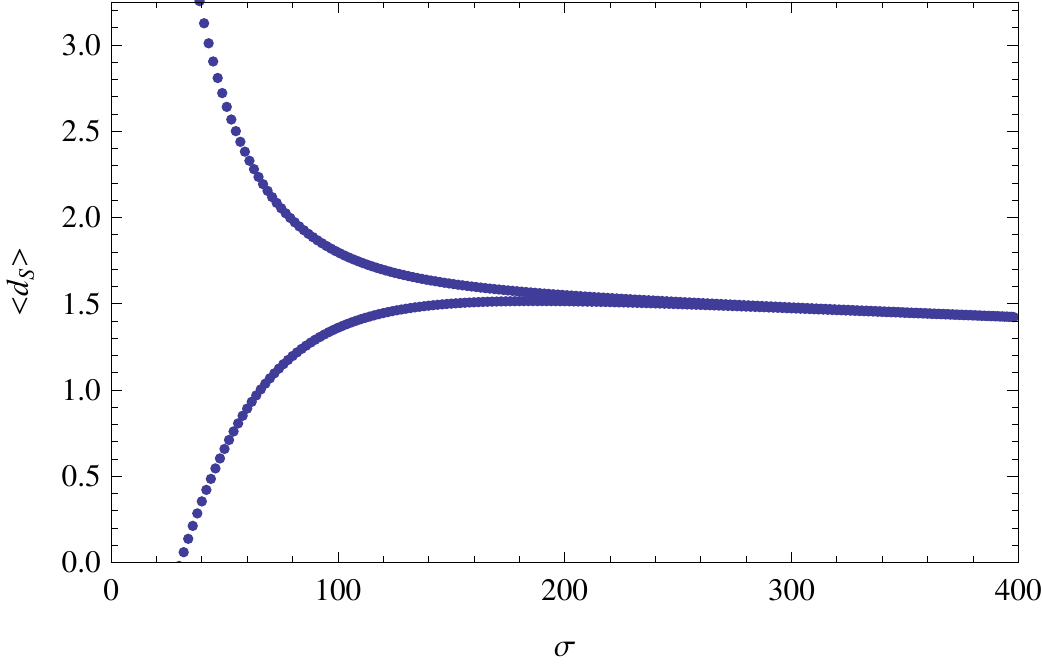}
\label{sd601000}
}
\subfigure[ ]{
\includegraphics[scale=0.49]{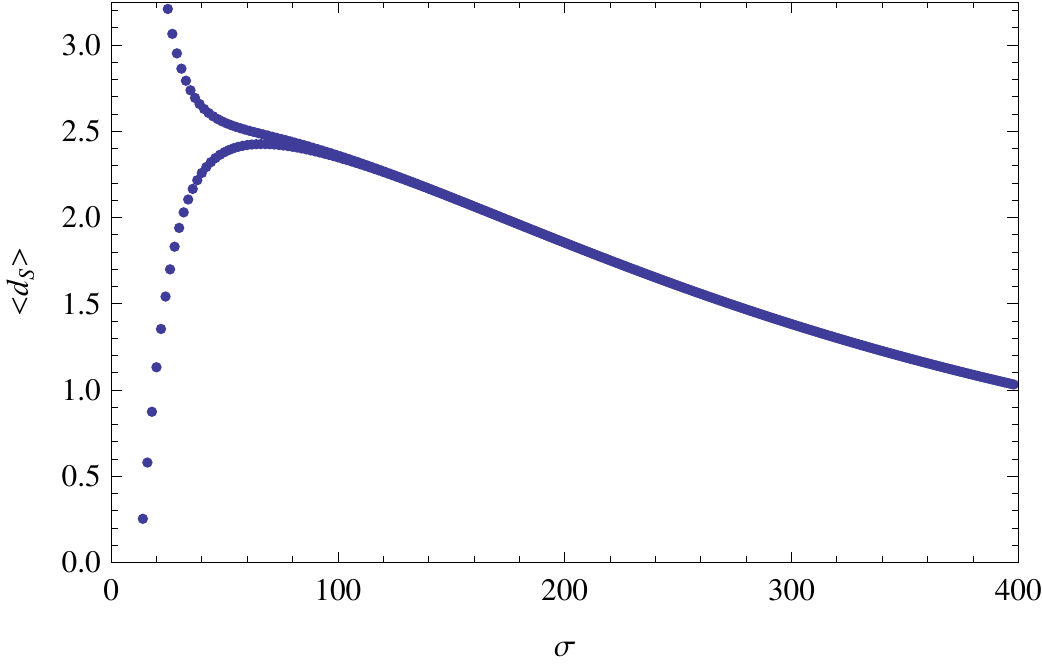}
\label{sd101000}
}
\label{CDTspectraldimension}
\caption[Optional caption for list of figures]{The ensemble average spectral dimension $\langle d_{s}\rangle$ as a function of diffusion time $\sigma$. \subref{sd601000} Phase A ($k_{0}=6.00$, $k_{3}=1.85$, $\lambda=1.00$, $\alpha=0.00$) \subref{sd101000} Phase C ($k_{0}=1.00$, $k_{3}=0.75$, $\lambda=1.00$, $\alpha=0.00$)}
\end{figure}

\subsection{Phase Diagram}

Our model's coupling constant space is $4$-dimensional. Based on the phase structures of both $(2+1)$- and $(3+1)$-dimensional causal dynamical triangulations, we expect that for fixed $N_{3}$ our model is only well-defined on a $3$-dimensional subspace. This critical surface approximates the so-called infinite volume surface, which corresponds to the limit in which $N_{3}$ increases without bound while the lattice spacing $a$ vanishes such that the product $N_{3}a^{3}$ remains constant. Supposing that our model possesses a second order phase transition, at which we could define its continuum limit, this transition must be located at a phase boundary on the critical surface.

Now, a $3$-dimensional subspace of largely unknown extent represents a formidably expansive space to explore numerically. Accordingly, we have limited our initial investigations to the subspace of the coupling constant space consisting of the $\lambda-\alpha$ plane near the origin for fixed $k_{0}$. Specifically, we set $k_{0}=1.00$, select values for $\lambda$ and $\alpha$, and then tune to the value of $k_{3}$ on the critical surface. With $\lambda=1.00$ and $\alpha=0.00$ this corresponds to a point in the physical phase of causal dynamical triangulations for the Regge action.

Within this region we have generated forty-seven ensembles of representative spacetimes, each for a different set $\{k_{0},\lambda,\alpha,k_{3}^{c}(k_{0},\lambda,\alpha)\}$ of the coupling constants. We display in figure $5$ the critical surface and the associated phase structure as ascertained thus far. Our explorations indicate the existence of three phases: a phase contiguous with the physical phase of causal dynamical triangulations for the Regge action that we call phase C;  a phase emerging for sufficiently large values of $\lambda$ that we call phase D; and a phase emerging for sufficiently large values of $\alpha$ that we call phase E. We are not entirely certain that phases D and E are distinct: our measurements might instead indicate modulation of a single phase across the relevant region in the $\lambda-\alpha$ plane, a possibility under active investigation. On the other hand, we are quite certain that phases D and E are not artifacts: our simulations exhibit the characteristic lengthening of the autocorrelation time near the phase boundaries, and the geometric properties of phases D and E persist under an increase in the total number of spacelike hypersurfaces. The presence of these novel phases of extended geometry counts as the first explicit indication that the spatial curvature squared terms in the action \eqref{eq:discHLE} exert a significant effect as opposed to being renormalized to irrelevance. This nicely matches the scaling behavior of the curvature squared terms expected from the analytic approach to Ho\v{r}ava-Lifshitz gravity.

\begin{figure}[ht]
\centering
\subfigure[ ]{
\includegraphics[scale=0.6]{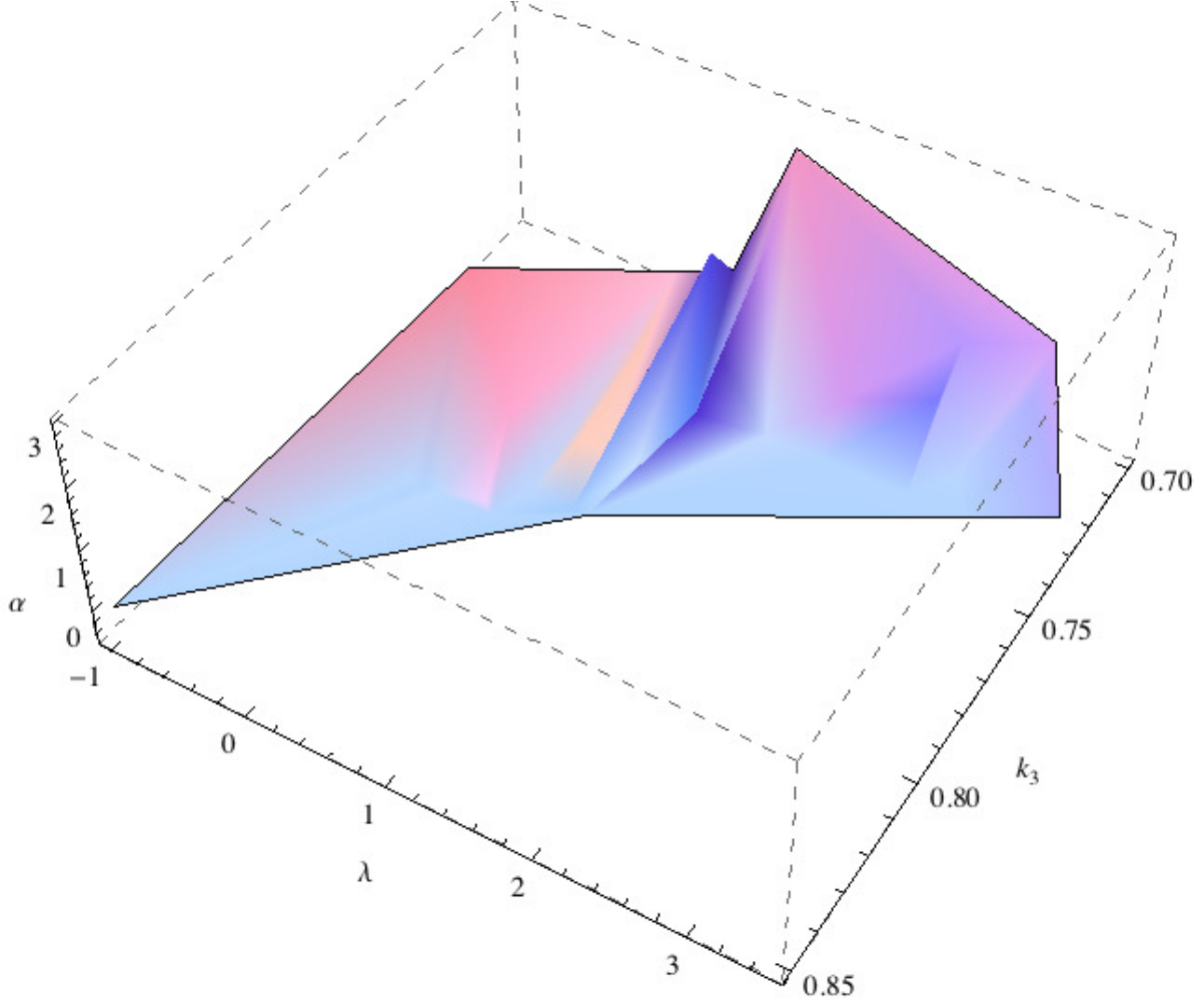}
\label{criticalsurface}
}
\subfigure[ ]{
\includegraphics[scale=0.35]{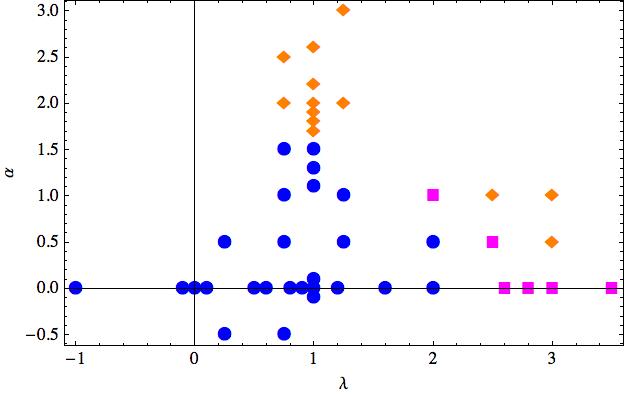}
\label{phasediagram}
}
\label{phases}
\caption{\subref{criticalsurface} The critical surface as explored thus far. \subref{phasediagram} The critical surface projected onto the $\lambda$-$\alpha$ plane showing phases C, D, and E respectively in blue circles, magenta squares, and orange diamonds.}
\end{figure}

In figure $6$ we depict spacetimes representative of each of the three phases. Specifically, we plot the number $N_{2}^{SL}$ of spacelike triangles---a measure of the discrete $2$-volume---as a function of discrete time $\tau$. Representative spacetimes in phase C are characterized by a single correlated accumulation of tetrahedra spread across a substantial range of $\tau$. Representative spacetimes in phase D are characterized by an intermittent series of accumulations of tetrahedra each spread across a small range of $\tau$. Representative spacetimes in the phase E are characterized by a moderately uniform distribution of tetrahedra spread across the entire range of $\tau$.

\begin{figure}[ht]
\centering
\subfigure[ ]{
\includegraphics[scale=0.455]{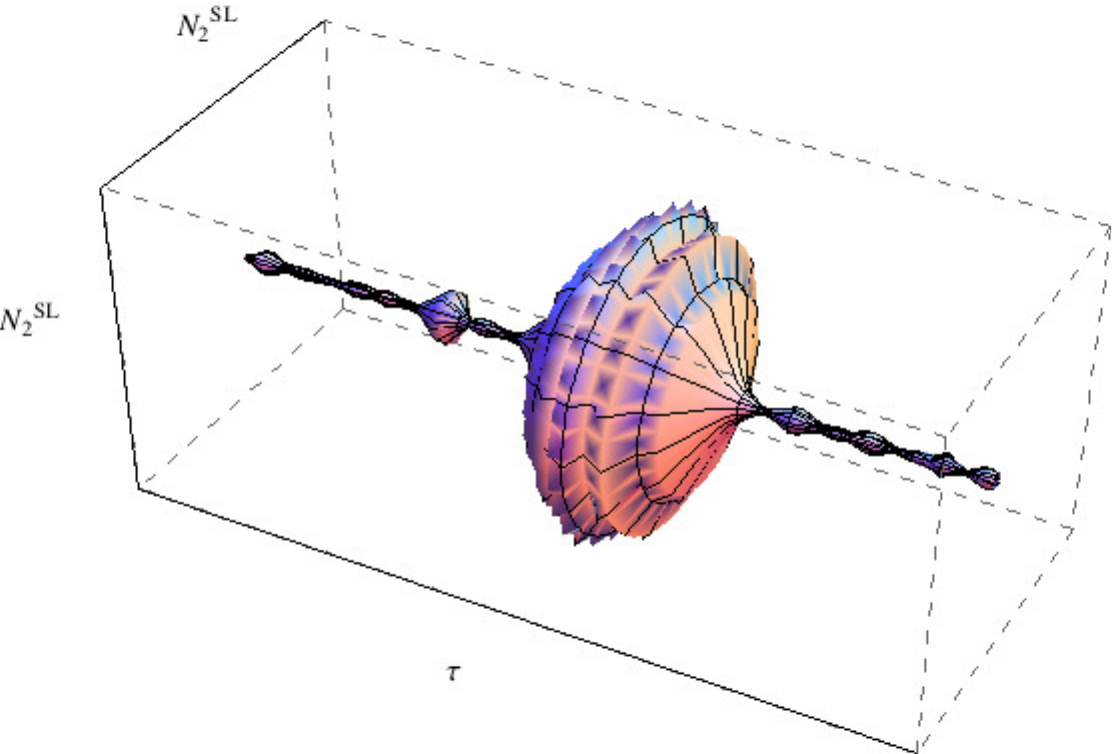}
\label{3d1007505}
}
\subfigure[ ]{
\includegraphics[scale=0.455]{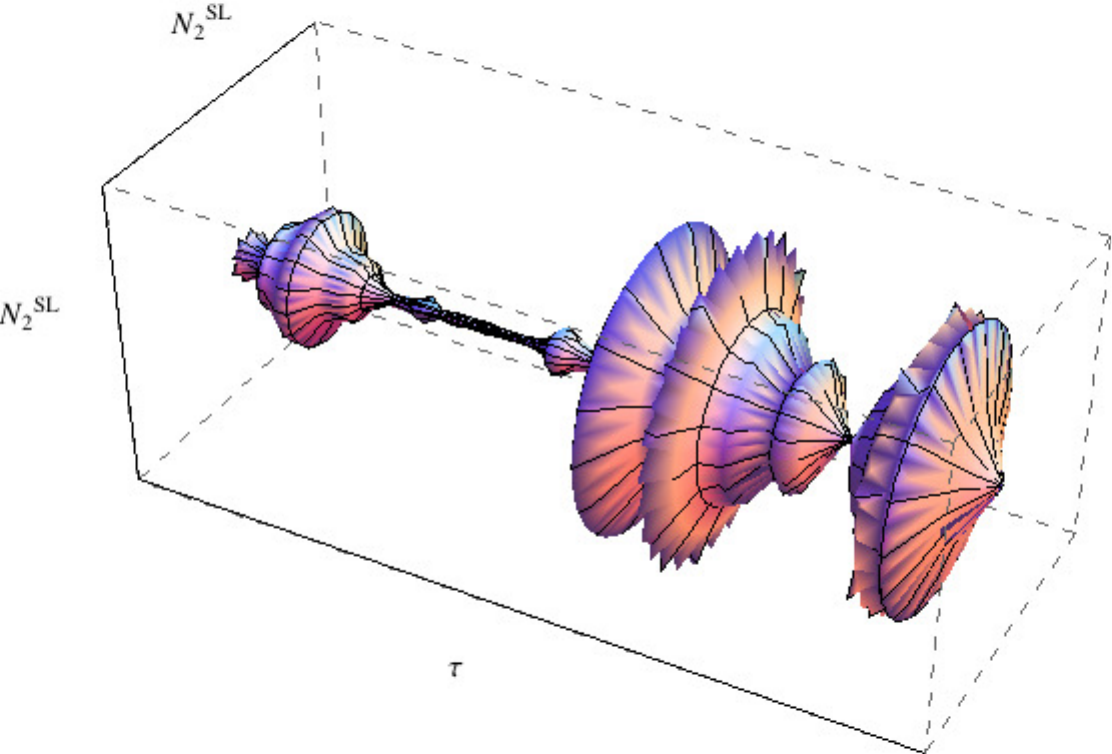}
\label{3d103500}
}
\subfigure[ ]{
\includegraphics[scale=0.455]{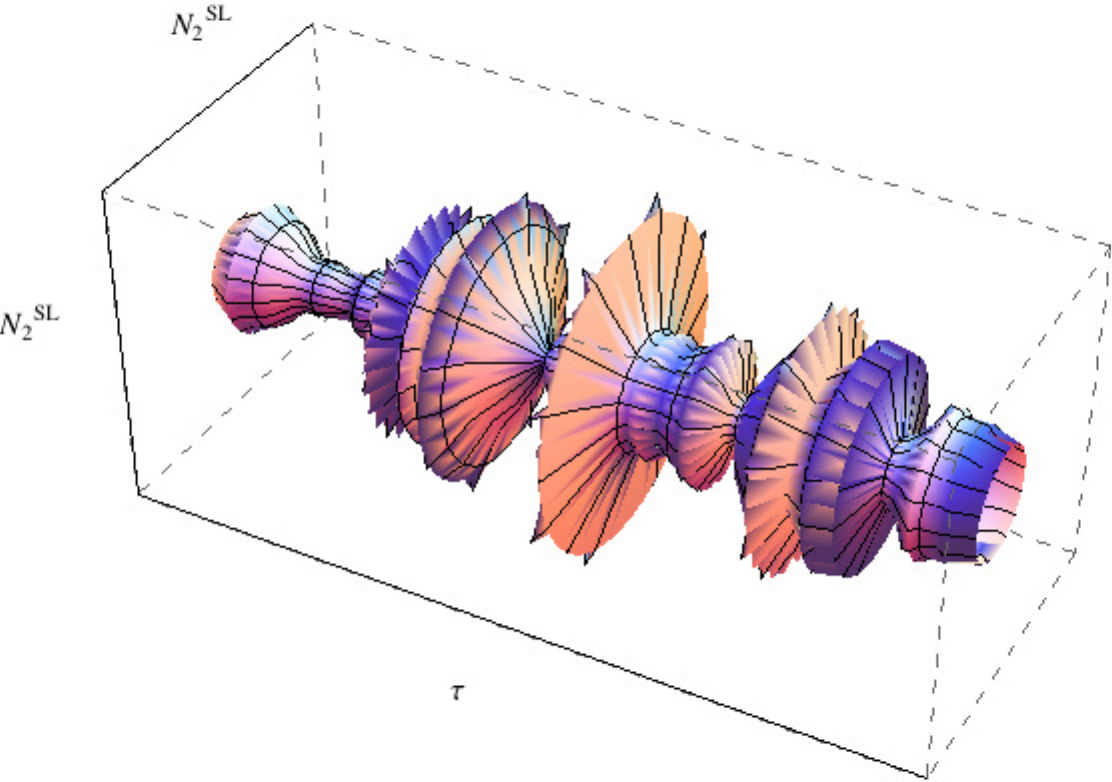}
\label{3d101026}
}
\subfigure{
\includegraphics[scale=0.49]{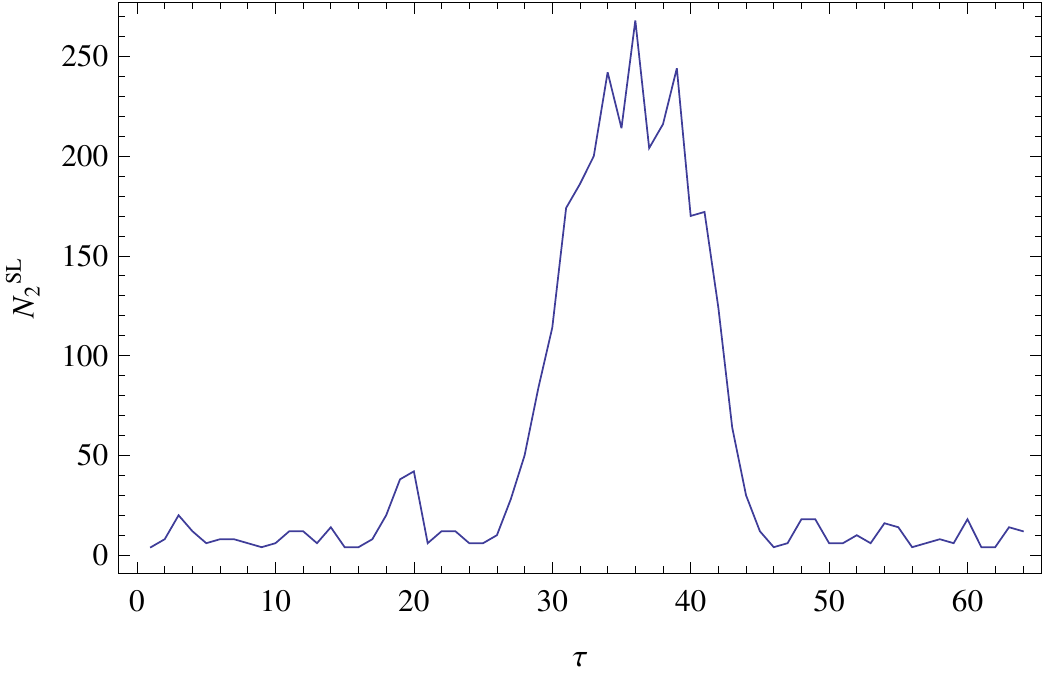}
\label{2d1007505}
}
\subfigure{
\includegraphics[scale=0.49]{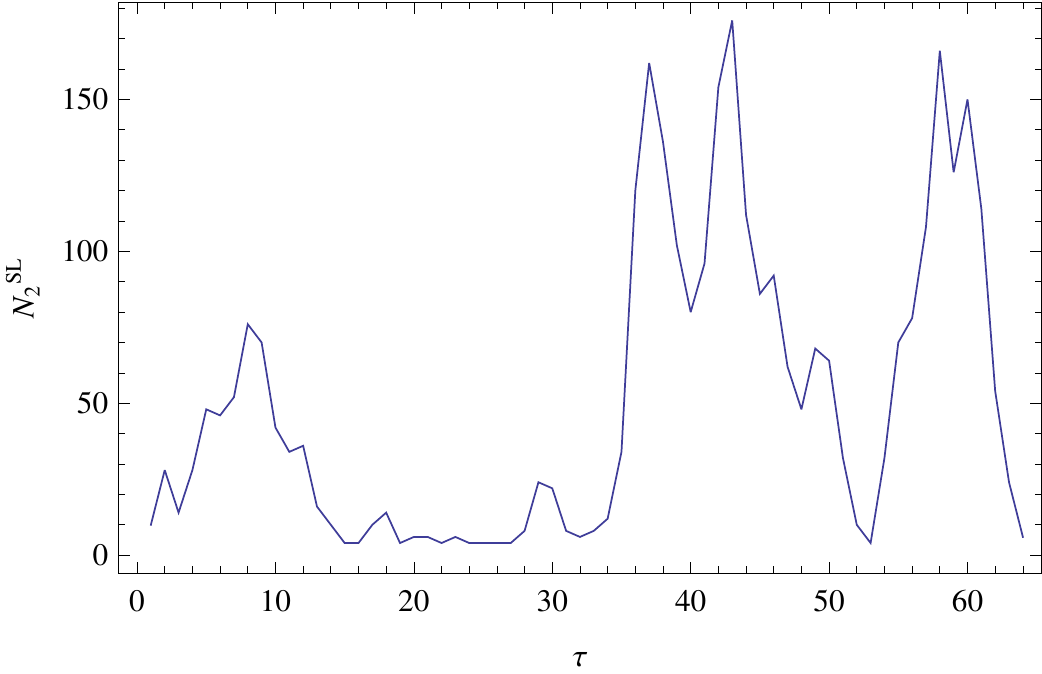}
\label{2d103500}
}
\subfigure{
\includegraphics[scale=0.49]{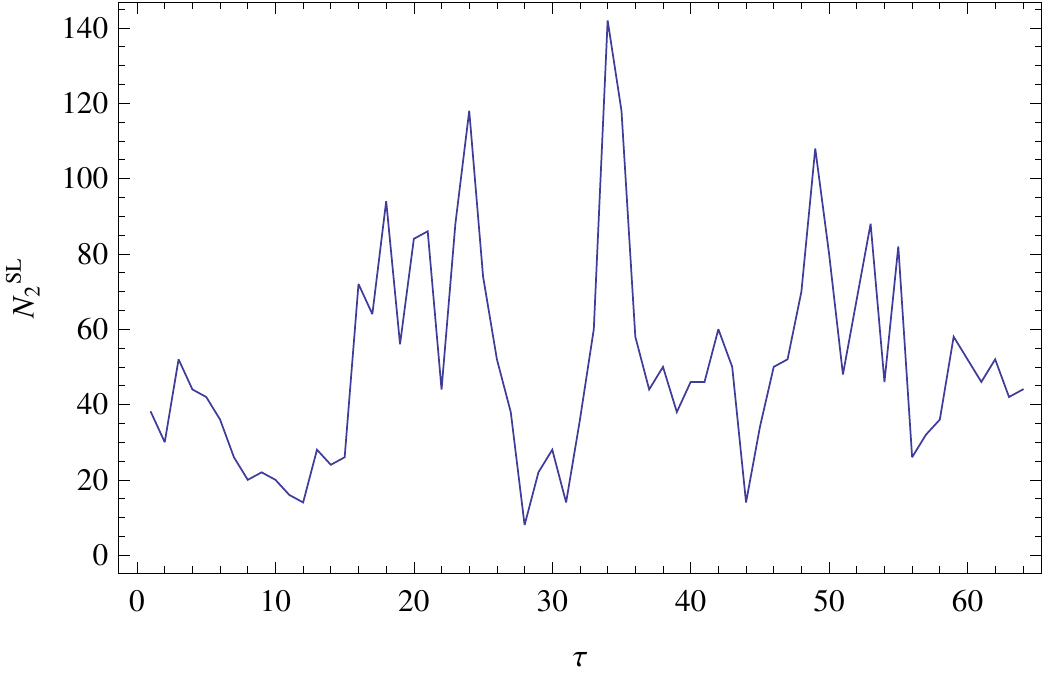}
\label{2d101026}
}
\label{HLrepspacetimes}
\caption[Optional caption for list of figures]{Depictions of representative spacetimes showing the number $N_{2}^{SL}$ of spacelike triangles as a function of discrete time $\tau$. \subref{3d1007505} Phase C ($k_{0}=1.00$, $k_{3}=0.77$, $\lambda=0.75$, $\alpha=0.50$) \subref{3d103500} Phase D ($k_{0}=1.00$, $k_{3}=0.73$, $\lambda=3.50$, $\alpha=0.00$) \subref{3d101026} Phase E ($k_{0}=1.00$, $k_{3}=0.72$, $\lambda=1.00$, $\alpha=2.60$)}
\end{figure}

\subsection{Physicality and Semiclassicality}\label{physsemi}

We now present preliminary evidence suggesting that the ensemble average geometries in all three phases show signs of being both physical and semiclassical. This evidence stems from analyses of two geometric observables associated with our ensembles of representative spacetimes: the spectral dimension as a function of diffusion time and the discrete $2$-volume as a function of discrete time. Based on these analyses, we draw certain comparisons to the relevant solutions of the classical theory now in imaginary time presented in appendix \ref{app1}.

\subsubsection{Spectral Dimension}

The spectral dimension of a space measures its effective dimensionality as experienced by a random walker. We determine the ensemble average spectral dimension by the method described, for instance, in  \cite{SpecDimHB}. Specifically, we directly measure the return probability $P_{r}(\sigma)$ as a function of diffusion time $\sigma$ for each representative spacetime in a given ensemble. We then compute the ensemble average spectral dimension as
\begin{equation}
\langle d_{s}(\sigma)\rangle=-2\frac{\mathrm{d}\ln{\langle P_{r}(\sigma)\rangle}}{\mathrm{d}\ln{\sigma}}
\end{equation}
employing an appropriate discretization of the derivative.

In figure $7$ we display plots of the ensemble average spectral dimension as a function of  diffusion time for each of the three phases. Before interpreting these plots, we should comment on the evident bifurcation for small values of diffusion time. This effect reflects  the discrete nature of our triangulated spacetimes: random walks of even and of odd lengths yield diverging estimates for the spectral dimension on scales sufficiently short in comparison to the discretization scale. Increasing the number of tetrahedra comprising each spacetime pushes the bifurcation scale towards smaller diffusion times, revealing the physical nature of the spectral dimension on such scales.

\begin{figure}[ht]
\centering
\subfigure[ ]{
\includegraphics[scale=0.49]{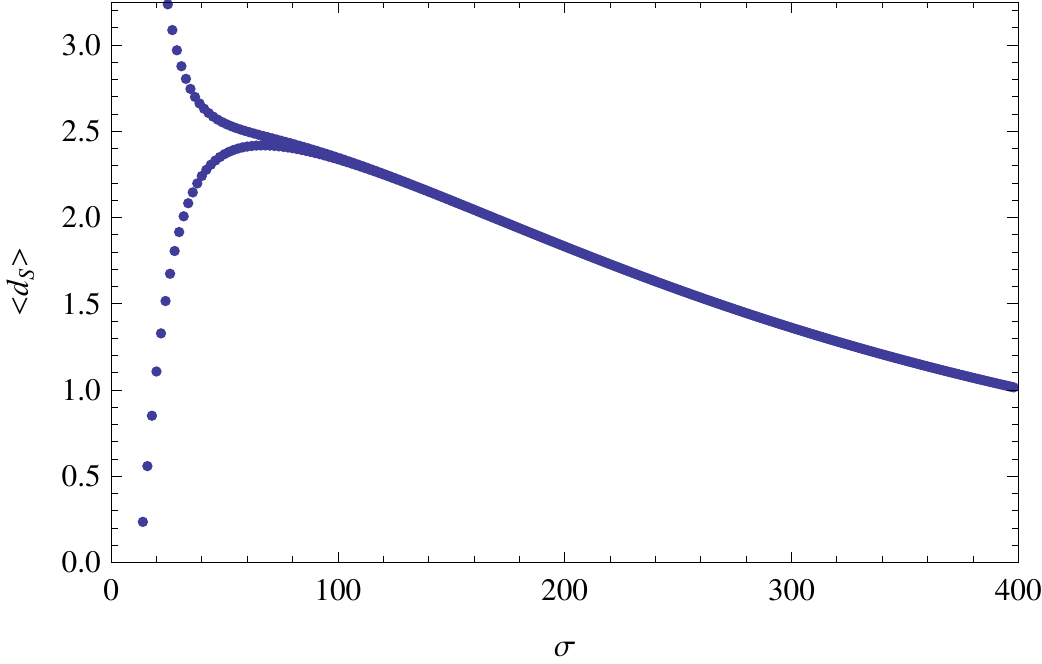}
\label{sd1007505}
}
\subfigure[ ]{
\includegraphics[scale=0.49]{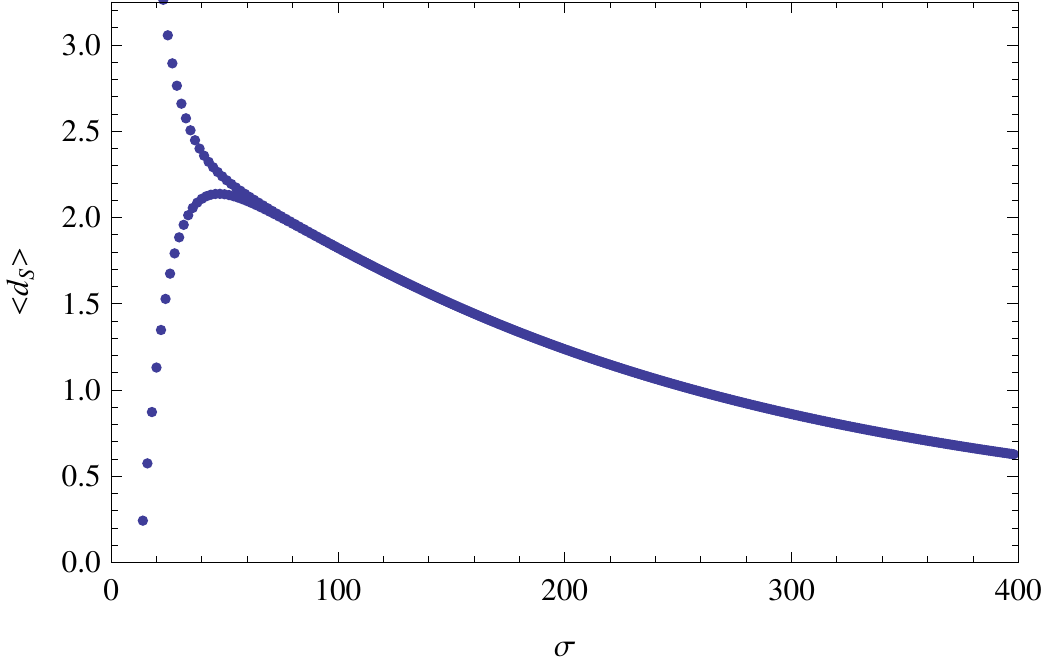}
\label{sd103500}
}
\subfigure[ ]{
\includegraphics[scale=0.49]{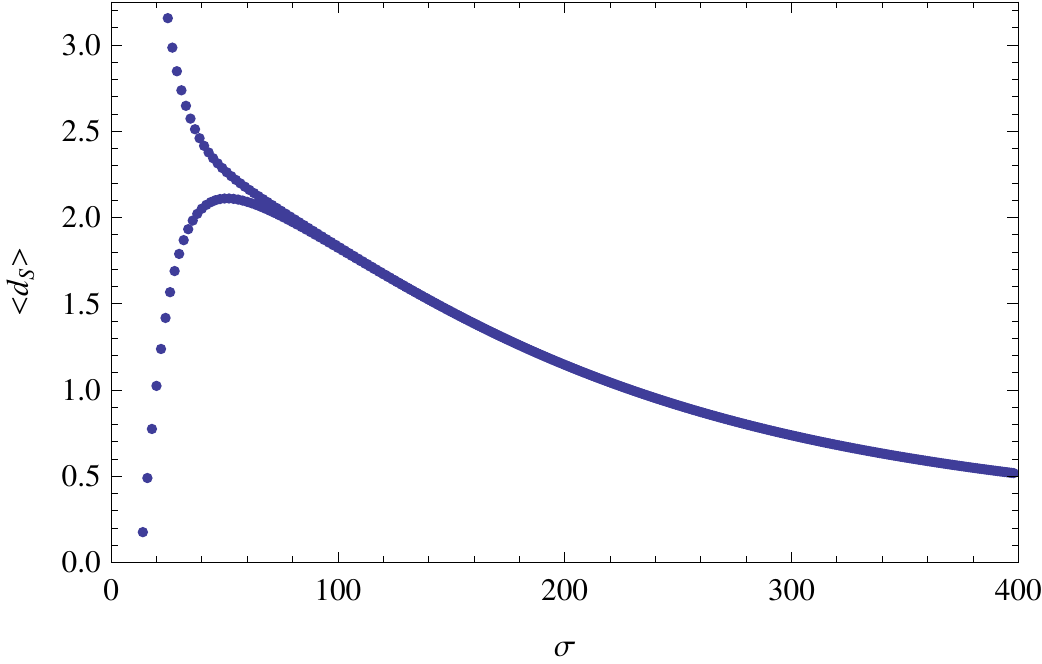}
\label{sd101026}
}
\label{HLspectraldimension}
\caption[Optional caption for list of figures]{The ensemble average spectral dimension $\langle d_{s}\rangle$ as a function of diffusion time $\sigma$. \subref{sd1007505} Phase C ($k_{0}=1.00$, $k_{3}=0.77$, $\lambda=0.75$, $\alpha=0.50$) \subref{sd103500} Phase D ($k_{0}=1.00$, $k_{3}=0.73$, $\lambda=3.50$, $\alpha=0.00$) \subref{sd101026} Phase E ($k_{0}=1.00$, $k_{3}=0.72$, $\lambda=1.00$, $\alpha=2.60$)}
\end{figure}

Now compare the plots of figure $7$ to those of ordinary causal dynamical triangulations in figure $4$. All three of the Ho\v{r}ava-Lifshitz spectral dimensions much more closely resemble that of figure $4$(b)---for the physical phase of causal dynamical triangulations---than that of figure $4$(a)---for the decoupled phase of causal dynamical triangulations. The resemblance between the plots of figures $7$(a) and $4$(b) is particularly close, as we might expect, since the two ensembles are both in phase C. In particular, the spectral dimensions shown in figure $7$ all peak at values between $2$ and $3$ for small diffusion times and then gradually decay for large diffusion times. We expect this decay at large diffusion times: these spacetimes are compact and have curvature.

For $(2+1)$-dimensional causal dynamical triangulations the spectral dimension in the physical phase reaches $3$ for ensembles characterized by larger values of $N_{3}$ \cite{SpecDimHB,JK1}. Such measurements provide a key piece of evidence for the semiclassical nature of the phase's ensemble average geometry. This suggests that the maxima of the spectral dimension plots in figures $4$(b) and $7$(a) are depressed by finite size effects. Assuming that such effects are comparable for our ensembles in phases D and E, we are led to conclude that the maxima of the spectral dimension in these phases is slightly below the topological value of $3$. In Ho\v{r}ava-Lifshitz gravity the spectral dimension is generally predicted to flow under renormalization, so potentially the plots of figure $7$ reflect such behavior \cite{PetrSpec}. We are currently running simulations in phases D and E for larger values of $N_{3}$ in the hope of resolving these issues. Nevertheless, we maintain that the similarities in form between the spectral dimensions depicted in figure $7$ and in figure $4$(b) hint at the physicality and semiclassicality of phases C, D, and E.

\subsubsection{Discrete $2$-Volume}

The foliated structure of causal triangulations allows for the measurement of certain quantities as functions of the discrete time coordinate. The number $N_{2}^{SL}$ of spacelike triangles on a given spacelike hypersurface is one such quantity. Although the value of $N_{2}^{SL}$ on a single spacelike hypersurface is not physically meaningful, the set $\{N_{2}^{SL}(\Sigma_{\tau})\}_{\tau=1}^{T}$ does contain physical information.

For an ensemble in phase C, there exist straightforward techniques for performing a coherent ensemble average of $\{N_{2}^{SL}(\Sigma_{\tau})\}_{\tau=1}^{T}$. (See, for instance, \cite{AJL6}.) These techniques rely on the characteristic feature of phase C spacetimes: the single accumulation of tetrahedra as depicted in figures $3$(b) and $6$(a). Intuitively, each method functions to align an appropriately defined center of this accumulation with the central value of the discrete time coordinate. Once accomplished for all of the representative spacetimes in an ensemble, the coherent ensemble average of $\{N_{2}^{SL}(\Sigma_{\tau})\}_{\tau=1}^{T}$ is defined by the discrete timewise average.

We employ a method that we call equal discrete $2$-volume splitting. The name refers to the algorithm for appropriately defining the center of the accumulation of tetrahedra. Formally, given a representative spacetime, form all divisions $D_{i}$ of that spacetime into two sets each of $\frac{T}{2}$ spacelike hypersurfaces, maintaining the periodic discrete time order of the spacelike hypersurfaces:
\begin{subequations}
\begin{eqnarray}
D_{1}&=&\left\{\left\{\Sigma_{1},\ldots,\Sigma_{\frac{T}{2}}\right\},\left\{\Sigma_{\frac{T}{2}+1},\ldots,\Sigma_{T}\right\}\right\}\\
D_{2}&=&\left\{\left\{\Sigma_{2},\ldots,\Sigma_{\frac{T}{2}+1}\right\},\left\{\Sigma_{\frac{T}{2}+2},\ldots,\Sigma_{1}\right\}\right\}\\
&\vdots&\nonumber\\
D_{T}&=&\left\{\left\{\Sigma_{T},\ldots,\Sigma_{\frac{T}{2}-1}\right\},\left\{\Sigma_{\frac{T}{2}},\ldots,\Sigma_{T-1}\right\}\right\}.
\end{eqnarray}
\end{subequations}
Then select the two particular divisions $\bar{D}_{eq}$ and $\tilde{D}_{eq}$ that most nearly equalize the discrete $2$-volume summed over the $\frac{T}{2}$ spacelike hypersurfaces in each set of the division: the quantity
\begin{equation}
\sum_{\tau\in\left\{i,\ldots,i+\frac{T}{2}-1\right\}}N_{2}^{SL}(\Sigma_{\tau})\quad-\,\sum_{\tau\in\left\{i+\frac{T}{2},\ldots,i-1\right\}}N_{2}^{SL}(\Sigma_{\tau})
\end{equation}
is minimized for both $\bar{D}_{eq}$ and $\tilde{D}_{eq}$. Note that $\bar{D}_{eq}$ and $\tilde{D}_{eq}$ only differ in the order of their two sets. Next relabel the discrete time coordinate over the set of values $\{-\frac{T}{2},-\frac{T}{2}+1,\ldots,\frac{T}{2}-1,\frac{T}{2}\}$ so that the values $\{-\frac{T}{2},\ldots,-\frac{1}{2}\}$ label the first set of spacelike hypersurfaces and the values $\{\frac{1}{2},\ldots,\frac{T}{2}\}$ label the second set of spacelike hypersurfaces in both $\bar{D}_{eq}$ and $\tilde{D}_{eq}$. Finally choose the division for which the values of $N_{2}^{SL}(\Sigma_{-\frac{1}{2}})$ and $N_{2}^{SL}(\Sigma_{\frac{1}{2}})$ are greatest. With all of the spacetimes in an ensemble so aligned, perform a discrete timewise average of $N_{2}^{SL}$ over the representative spacetimes.

In figure $8$ we show the results of the equal discrete $2$-volume splitting average for two ensembles in phase C. The data points indicate the equal discrete $2$-volume splitting average value of $\{N_{2}^{SL}(\Sigma_{\tau})\}_{\tau=1}^{T}$; the light vertical bars indicate one standard deviation of error. The thin curve is a one parameter fit of the data points within the central accumulation of tetrahedra to the functional form
\begin{equation}
N_{2}^{SL}(\tau)=\frac{2}{\pi}\frac{\langle N_{3}^{(3,1)}\rangle}{\tilde{s}_{0}\langle N_{3}^{(1,3)}\rangle^{1/3}}\cos^{2}{\left(\frac{\tau}{\tilde{s}_{0}\langle N_{3}^{(1,3)}\rangle^{1/3}}\right)},
\end{equation}
which is a discretization of the $2$-volume as a function of global time of Euclidean de Sitter spacetime. (See \cite{JK2} and \cite{AJL6} for a derivation of this discretization in $2+1$ and $3+1$ dimensions, respectively.) Substantial evidence already demonstrates that, for the general relativistic values of $\lambda$ and $\alpha$, the ensemble average geometry in phase C closely approximates Euclidean de Sitter spacetime \cite{Semiclassical,3Dnonpert,AJL2,AJL3,AJL4,SpecDimAmbjorn,AJL5,AJL6,SpecDimHB,JK2}. The plot in figure $8$(b) provides the first evidence that the ensemble average geometry also has this property for the broader range of nonrelativistic values of both $\lambda$ and $\alpha$.

\begin{figure}[ht]
\centering
\subfigure[ ]{
\includegraphics[scale=0.49]{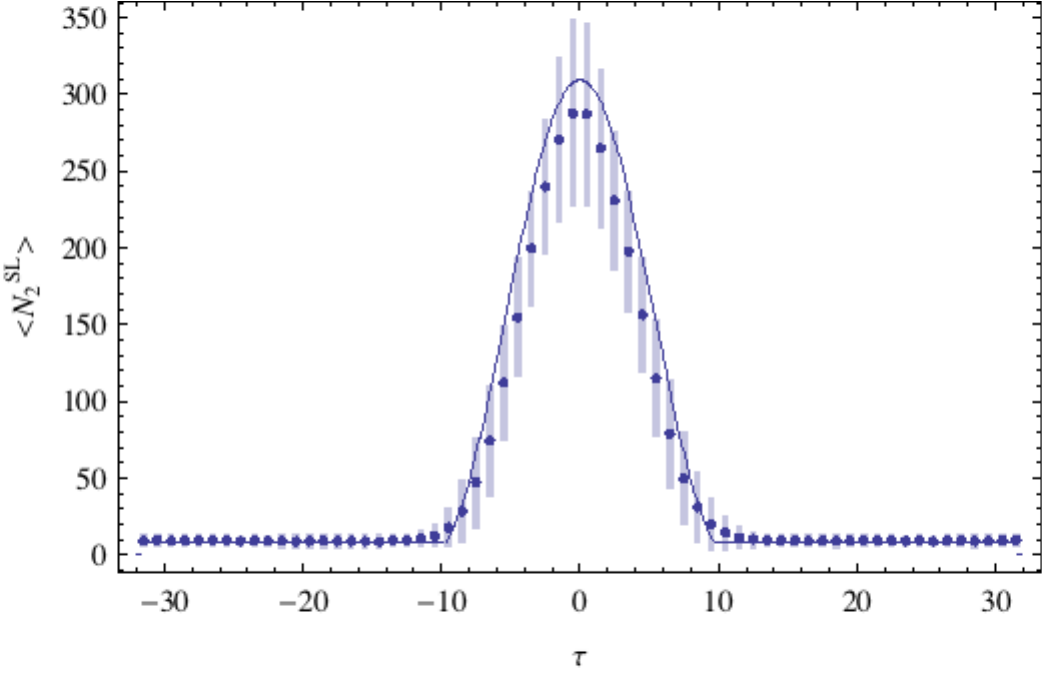}
\label{volprof101000}
}
\subfigure[ ]{
\includegraphics[scale=0.49]{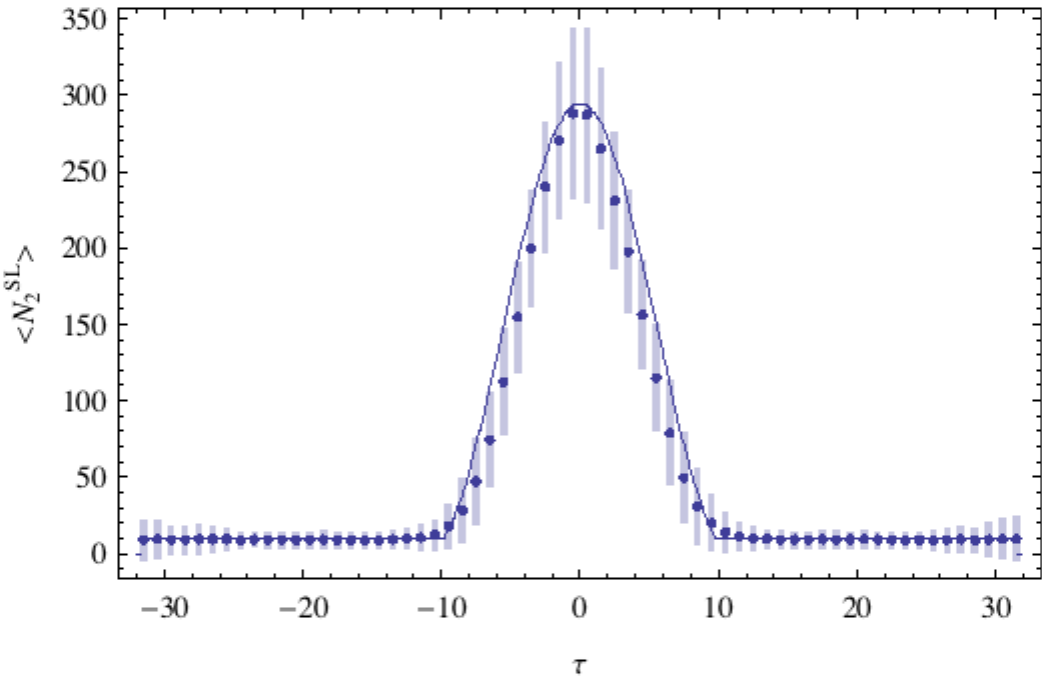}
\label{volprof1007505}
}
\label{volumeprofile}
\caption[Optional caption for list of figures]{The ensemble average number $\langle N_{2}^{SL}\rangle$ of spacelike triangles as a function of discrete time $\tau$. \subref{volprof101000} Phase C ($k_{0}=1.00$, $k_{3}=0.75$, $\lambda=1.00$, $\alpha=0.00$) with $\tilde{s}_{0}=0.46$ \subref{volprof1007505} Phase C ($k_{0}=1.00$, $k_{3}=0.77$, $\lambda=0.75$, $\alpha=0.50$) with $\tilde{s}_{0}=0.48$}
\end{figure}

Ho\v{r}ava-Lifshitz gravity admits Euclidean de Sitter spacetimes: for $\lambda>\frac{1}{2}$ and $\alpha=0$, the solutions \eqref{eq:solelongated} and \eqref{eq:soldeSitter} coincide in their description of such spacetimes. Note, however, that for relatively small values of $\alpha$, the solutions \eqref{eq:solelongated} and \eqref{eq:soldeSitter} do not deviate too markedly from Euclidean de Sitter spacetime. Given the presence of finite size effects and the inherent error in our measurements, one of these solutions for $\alpha\neq 0$ may well fit better the ensemble average geometry in phase C. Our current data are not sufficiently detailed for us to make a conclusive statement. Furthermore, there exists some evidence \cite{SpecDimHB} that the ensemble average geometry in $(2+1)$-dimensional causal dynamical triangulations is a deformed version of Euclidean de Sitter spacetime. Supposing that this is the case, the process of taking the continuum limit must generate some additional terms besides those in the Regge action.

The two plots in figure $8$ quite closely resemble one another---in both spatial and temporal extents of the central accumulation of discrete $2$-volume---even though that of figure $8$(a) is for general relativistic values of $\lambda$ and $\alpha$ while that of figure $8$(b) is for nonrelativistic values of $\lambda$ and $\alpha$. To dispel the suspicion that the $K^{2}$ and $R_{2}^{2}$ terms are renormalized to irrelevance in phase C, we present in figure $9$ four plots of the equal discrete $2$-volume splitting average of $\{N_{2}^{SL}(\Sigma_{\tau})\}_{\tau=1}^{T}$ each for a different value of $\lambda$. Clearly, varying $\lambda$ affects the ensemble average geometry that emerges.

\begin{figure}
\centering
\includegraphics[scale=0.49]{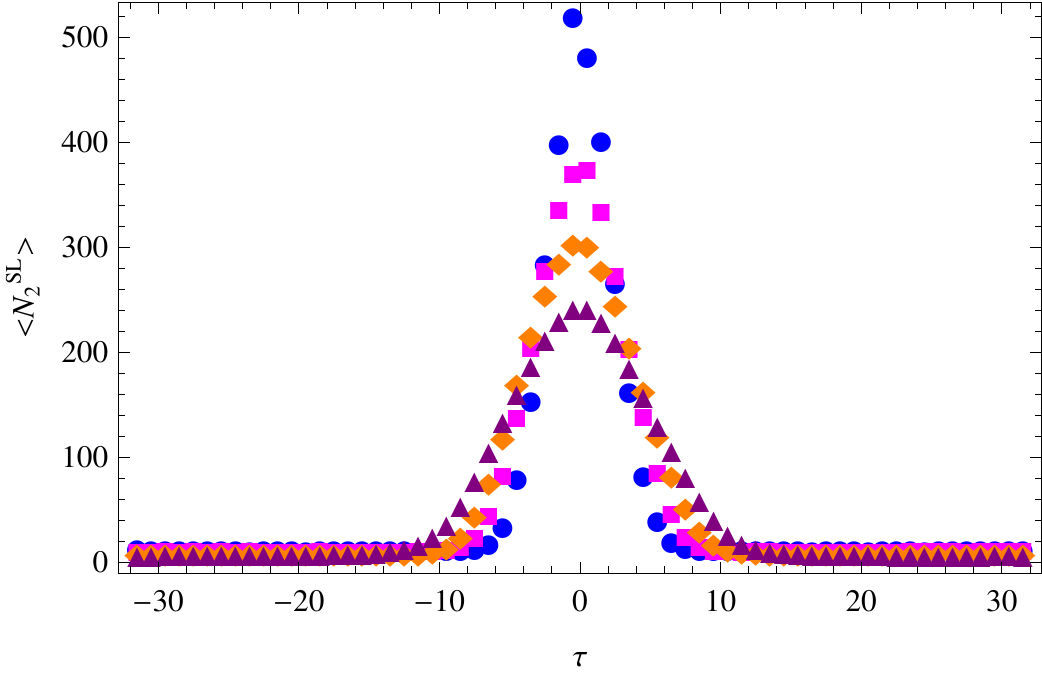}
\label{lambdavar}
\caption{The ensemble average number $\langle N_{2}^{SL}\rangle$ of spacelike triangles as a function of discrete time $\tau$ for four ensembles in phase C differing in their respective values of the coupling constant $\lambda$: blue circles ($k_{0}=1.00$, $k_{3}=0.85$, $\lambda=-1.00$, $\alpha=0.00$), magenta squares ($k_{0}=1.00$, $k_{3}=0.79$, $\lambda=0.00$, $\alpha=0.00$), orange diamonds ($k_{0}=1.00$, $k_{3}=0.75$, $\lambda=1.00$, $\alpha=0.00$), purple triangles ($k_{0}=1.00$, $k_{3}=0.74$, $\lambda=2.00$, $\alpha=0.00$).}
\end{figure}

For phases D and E we currently do not know how to coherently average $\{N_{2}^{SL}(\Sigma_{\tau})\}_{\tau=1}^{T}$ over an ensemble. In an effort to determine a method, we computed the ensemble average power in the discrete Fourier transform of $\{N_{2}^{SL}(\Sigma_{\tau})\}_{\tau=1}^{T}$, the results of which we display in figure $10$. As these plots show, there is no notable periodicity present in these ensemble's average geometry since virtually all of the power falls in the zero frequency mode. This lack of periodicity may in fact be indicative of the semiclassical nature of phase E. With $\alpha>0$ the classical equations of motion \eqref{eq:momconstraint} and \eqref{eq:metriceom} are also satisfied for the \emph{Ansatz} \eqref{ansatz} when the squared scale factor $a^{2}(t)$ has the constant value $\sqrt{\frac{\alpha}{2\Lambda}}$. The depiction in figure $6$(c) of a representative spacetime in phase E resembles to a certain extent a spacetime with approximately constant scale factor, and the plot of figure $10$(b) reinforces this interpretation. Moreover, this accords with our expectation that as currently devised the Markov chain Monte Carlo simulations converge on spacetimes globally minimizing the action \eqref{eq:discHLE}.

\begin{figure}[ht]
\centering
\subfigure[ ]{
\includegraphics[scale=0.49]{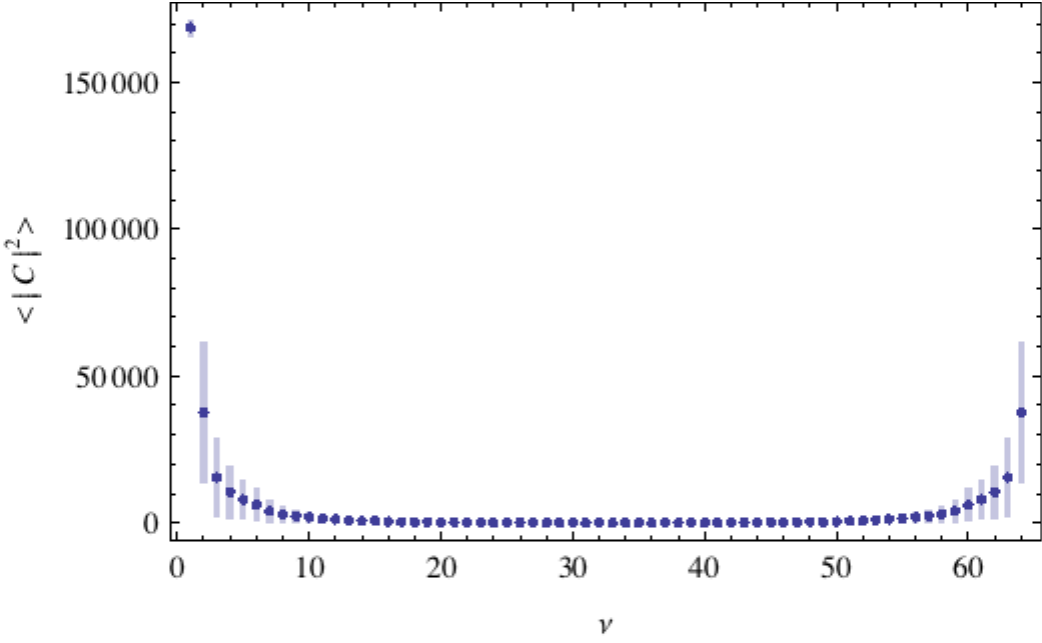}
\label{avpower103500}
}
\subfigure[ ]{
\includegraphics[scale=0.49]{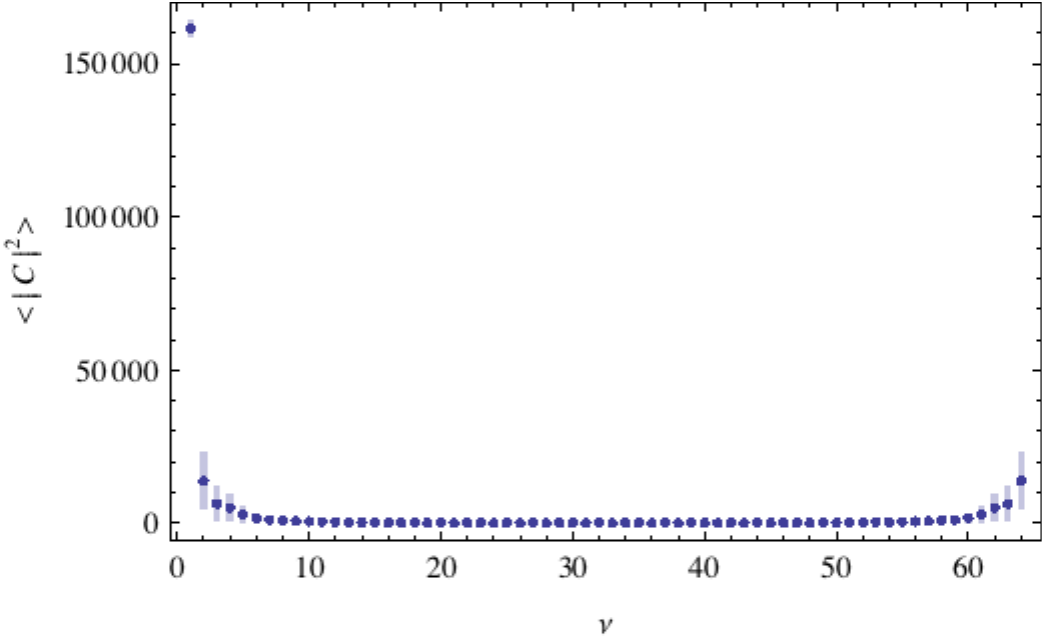}
\label{avpower101026}
}
\label{avpower}
\caption[Optional caption for list of figures]{The ensemble average power $\langle |c|^{2}\rangle$ as a function of discrete frequency $\nu$ in the discrete Fourier transform of the number $N_{2}^{SL}$ of spacelike triangles as a function of discrete time $\tau$.  \subref{avpower103500} Phase D ($k_{0}=1.00$, $k_{3}=0.73$, $\lambda=3.50$, $\alpha=0.00$) \subref{avpower101026} Phase E ($k_{0}=1.00$, $k_{3}=0.72$, $\lambda=1.00$, $\alpha=2.60$)}
\end{figure}

At first glance, however, the deviations from constant scale factor as illustrated in figure $6$(c) do not appear particularly small. To make a proper judgment, we compare the deviations in $\{N_{2}^{SL}(\Sigma_{\tau})\}_{\tau=1}^{T}$ between the ensemble in phase C associated with figure $6$(a) and the ensemble in phase E associated with figure $6$(c). Specifically, for each representative spacetime of this ensemble in phase E, we calculate the discrete time averaged deviation $\Delta_{N_{2}^{SL}}$ of the discrete $2$-volume from the mean. The ensemble average $\langle\Delta_{N_{2}^{SL}}\rangle$ provides a reasonable measure of the deviations from a time-independent geometry. For this ensemble in phase C, we calculate the deviations $\langle\Delta_{N_{2}^{SL}}\rangle_{min}$ and $\langle\Delta_{N_{2}^{SL}}\rangle_{max}$ of the discrete $2$-volume from the mean for the spacelike hypersurfaces of minimal and of maximal $\langle N_{2}^{SL}\rangle$ within the central accumulation of tetrahedra. The values $\langle\Delta_{N_{2}^{SL}}\rangle_{min}$ and $\langle\Delta_{N_{2}^{SL}}\rangle_{max}$ provide a reasonable measure of the range of deviations from a Euclidean de Sitter geometry. We find that
\begin{equation}
\frac{\langle\Delta_{N_{2}^{SL}}\rangle_{max}}{\langle N_{2}^{SL}\rangle_{max}}\Bigg|_{C}=0.20\quad<\quad\frac{\langle\Delta_{N_{2}^{SL}}\rangle}{\langle N_{2}^{SL}\rangle}\Bigg|_{E}=0.69\quad<\quad\frac{\langle\Delta_{N_{2}^{SL}}\rangle_{min}}{\langle N_{2}^{SL}\rangle_{min}}\Bigg|_{C}=0.78
\end{equation}
and that
\begin{equation}
\frac{\sqrt{\langle\Delta_{N_{2}^{SL}}\rangle_{max}}}{\sqrt[3]{\langle N_{3}\rangle}}\Bigg|_{C}=0.35\quad>\quad\frac{\sqrt{\langle\Delta_{N_{2}^{SL}}\rangle}}{\sqrt[3]{\langle N_{3}\rangle}}\Bigg|_{E}=0.27\quad>\quad\frac{\sqrt{\langle\Delta_{N_{2}^{SL}}\rangle_{min}}}{\sqrt[3]{\langle N_{3}\rangle_{min}}}\Bigg|_{C}=0.22.
\end{equation}
To achieve a more faithful comparison, we have considered the above two ratios instead of the deviations themselves. Although not definitive, the fact that the ratios for phase E fall between those for phase C lends credence to our interpretation of the semiclassical nature of phase E.

Supposing that further analysis supports the conjecture that time-independent geometries dominate phase E, how could one then interpret the C-E phase transition? A possible answer involves the global gravitational Hamiltonian $\mathcal{H}_\perp$ of \eqref{hperp}. Recall that, in our implementation of Ho\v{r}ava-Lifshitz gravity into the framework of causal dynamical triangulations,  we chose not to impose the condition \eqref{eq:intconstraint} as a constraint, instead treating $\mathcal{H}_\perp$ as a conserved global charge, essentially the total energy. In our statistical ensemble whether $\mathcal{H}_\perp$ vanishes or has a nontrivial spectrum on physical states thus becomes a question of dynamics. If the dynamics enforce the vanishing of $\mathcal{H}_\perp$ on physical states, then we might interpret this phenomenon as dynamical confinement of the gravitational charge.  A similar phenomenon has recently been discovered in $(2+1)$-dimensional relativistic chiral gravity \cite{alexetal}. There not just one gravitational charge but an infinite hierarchy of conserved chiral charges are confined, that is, dynamically vanish on all physical states of finite energy.

Now consider the hypothetical phase in which $\mathcal{H}_\perp$ is dynamically confined.  On spacetime geometries of the Friedmann-Lema\^\i tre-Robertson-Walker type, the Hamiltonian constraint equation becomes the Friedmann equation for the scale factor. The Friedmann equation  precludes the ground state geometry from being time-independent, forcing a cosmological, de Sitter-like evolution of the universe. In the context of causal dynamical triangulations, we observe this behavior in phase C.

In the hypothetical deconfined phase, on the other hand, the situation is more reminiscent of a typical condensed matter system:  the total Hamiltonian simply measures the system's energy levels with the ground state identified as the lowest energy state, typically static. We apparently observe this behaviour in our phase E.  Accordingly, we speculate that the C-E phase transition may be viewed as the deconfinement of the gravitational charge $\mathcal{H}_\perp$.

\section{Outlook}\label{outlook}

Motivated by a suite of striking similarities between Ho\v{r}ava-Lifshitz gravity and causal dynamical triangulations, we constructed a discrete version of the action for $(2+1)$-dimensional Ho\v{r}ava-Lifshitz gravity adapted to the formalism of causal dynamical triangulations. Using this action in Markov chain Monte Carlo simulations of the corresponding path integral, we found significant evidence for the existence of extended phases of geometry approximating certain classical solutions of Ho\v{r}ava-Lifshitz gravity. Quantum Ho\v{r}ava-Lifshitz gravity thus appears amenable to and compatible with the techniques of causal dynamical triangulations.

Since we have only just initiated this study, the prospects for further research are expansive. In the near term we have three primary goals for ongoing research. First, we plan to map more extensively our model's coupling constant space to determine the extent of the phases so far discovered and to identify the locations of phases not yet discovered. Specifically, we hope to ascertain how the curvature squared terms affect the decoupled phase of causal dynamical triangulations. Second, we wish to further our analysis of the semiclassical natures of phases C, D, and E along the lines of  \cite{Semiclassical,AJL6,SpecDimHB,JK2}. Third, we want to establish the orders of our model's phase transitions. In this direction we have attempted to determine order parameters for the C-D and C-E phase transitions; unfortunately, none of our trial order parameters have yet distinguished between these adjacent phases.  If our speculation about the confinement-deconfinement nature of the C-E phase transition is correct, then the appropriately defined ground state energy may act as an order parameter.

In the long term the richness of the literature on Ho\v{r}ava-Lifshitz gravity provides a host of directions for continuing research. First of all, recall that the simplest versions of $(2+1)$-dimensional Ho\v{r}ava-Lifshitz gravity, both projectable and nonprojectable, possess a single propagating scalar degree of freedom.  This is in stark contrast to the case of not only $(2+1)$-dimensional general relativity, but also the nonrelativistic generally covariant version of Ho\v{r}ava-Lifshitz gravity constructed in \cite{GenCovarC}. This scalar mode has generated considerable controversy with regard to its phenomenological viability \cite{Sot,Visser}, but its full dynamics, especially around the stable ground state, remains poorly understood. The $(2+1)$-dimensional theory should provide a clear window into the scalar mode's dynamics since the complications of propagating tensor modes are absent. By studying the nonperturbative dynamics of the scalar mode using causal dynamical triangulations, we hope to shed light on this issue. Our first challenge is the identification of an observable in causal dynamical triangulations that measures the number of local propagating degrees of freedom or that at least distinguishes between the absence and presence of local propagating degrees of freedom. Given the relative ease of extracting correlation functions of observables associated with the spacelike hypersurfaces of causal triangulations, we are currently working to understand the behavior of our model's conformal mode in relation to the scalar mode. We hope that such an investigation might illuminate the nature of the scalar dynamics.

Relatedly, perturbations of projectable Ho\v{r}ava-Lifshitz gravity about $(3+1)$-dimensional Minkowski spacetime generate instabilities \cite{Pathological}. This finding is of course not surprising: Minkowski spacetime is not the ground state of the commonly analyzed models. The situation is much improved on the background of de Sitter spacetime \cite{robertetal,Wang}. As the coupling constant $\lambda$ approaches unity, however, the higher derivative terms become relevant, leading to a breakdown of the linearized analysis. The authors of  \cite{Cosmo,Wang} have suggested that a nonperturbative Vainshtein mechanism might take effect, rendering this limit continuous to the general relativistic values of the coupling constants. Potentially, we could uncover this transition behavior in Markov chain Monte Carlo simulations. Indeed, the critical surface depicted in figure $5$(a) provides a small but intriguing piece of evidence for that possibility: the apparent geometric feature along the $\lambda=1$ direction.

To assess the phenomenological viability of at least the simple version of Ho\v{r}ava-Lifshitz gravity on which we have focused our study, we need to study the renormalization group flow of the coupling constants for the purpose of comparing the long distance behavior to that of general relativity. In \cite{Coarse} Henson proposes a coarse graining procedure for causal dynamical triangulations on which one could try to base a renormalization group procedure. If this scheme proves apt -- a question that we are currently exploring -- then we plan to employ the procedure to study the renormalization group flows of our model. Ultimately, we would like to explore the general conjecture formulated by several groups \cite{CDTandHL,GenCovar} suggesting that Ho\v{r}ava-Lifshitz gravity and causal dynamical triangulations belong to the same universality class. This circumstance would neatly explain the remarkable resemblances between these two approaches to the quantization of gravity.

\section*{Acknowledgments}

We  wish to thank Jan Ambj\o rn, Dario Benedetti, Diego Blas, Ted Jacobson, Renate Loll, Charles Melby-Thompson, Oriol Pujol\`as, Kevin Schaeffer, Sergey Sibiryakov, Thomas Sotiriou, Lewis Tunstall Garcia-Huidobro, Matt Visser, and Silke Weinfurtner for illuminating discussions at various stages of this work. C. Anderson acknowledges support from National Science Foundation under REU grant PHY-1004848 at the University of California, Davis. S. J. Carlip, J. H. Cooperman, and R. K. Kommu acknowledge support from Department of Energy under grant DE-FG02-91ER40674. P. Ho\v{r}ava and P. R. Zulkowski acknowledge support from National Science Foundation under Grant PHY-0855653, Department of Energy under Grant DE-AC02-05CH11231, and the Berkeley Center for Theoretical Physics.

\appendix

\section{On the Classical Equations of Motion}\label{app1}

We collect here the classical equations of motion stemming from the action \eqref{HLaction} for constant lapse and their solutions relevant to the Markov chain Monte Carlo simulations reported above. In Riemannian signature and for spacelike hypersurfaces having the topology of $ \mathcal{S}^{2} $, the equations of motion are
\begin{eqnarray}
0=\nabla_{i} \pi^{ij}(t,\mathbf{x})\label{eq:momconstraint}
\end{eqnarray}
and
\begin{eqnarray}
0&=&-\frac{1}{\sqrt{\gamma(t,\mathbf{x})}} \partial_{t} \left[  \sqrt{\gamma(t,\mathbf{x})} \pi^{ij}(t,\mathbf{x}) \right]+\frac{1}{2} \gamma^{ij}(t,\mathbf{x}) \left[K_{kl}(t,\mathbf{x})K^{kl}(t,\mathbf{x})-\lambda K^{2}(t,\mathbf{x}) -\alpha R_{2}^{2}(t,\mathbf{x}) +2 \Lambda \right] \nonumber \\ && -2 K^{il}(t,\mathbf{x})K_{l}^{j}(t,\mathbf{x})+ 2 \lambda K(t,\mathbf{x}) K^{ij}(t,\mathbf{x}) +2 \alpha \nabla^{i} \nabla^{j} R_{2}(t,\mathbf{x})- 2 \alpha \nabla^{2} R_{2}(t,\mathbf{x}) \gamma^{ij}(t,\mathbf{x}) \nonumber \\ && +\frac{1}{2} \left[ \nabla_{l} N^{i}(t,\mathbf{x}) \pi^{jl}(t,\mathbf{x}) + \nabla_{l} N^{j}(t,\mathbf{x}) \pi^{il}(t,\mathbf{x}) - \pi^{ij}(t,\mathbf{x}) \nabla_{l}N^{l}(t,\mathbf{x}) \right]\label{eq:metriceom}
\end{eqnarray}
where
\begin{equation}
\pi_{ij}(t,\mathbf{x}) = K_{ij}(t,\mathbf{x})-\lambda K(t,\mathbf{x}) \gamma_{ij}(t,\mathbf{x}).
\end{equation}
The momentum constraint \eqref{eq:momconstraint} results from variation of the shift vector, and the metric equations of motion \eqref{eq:metriceom} result from variation of the spatial metric tensor. Again we exclude the nonlocal integral constraint \eqref{eq:intconstraint} arising from variation of the lapse.

We seek solutions to the equations of motion \eqref{eq:momconstraint} and \eqref{eq:metriceom} in the form of the Friedmann-Lema\^\i tre-Robertson-Walker \emph{Ansatz}
\begin{equation}\label{ansatz}
\gamma(t,\mathbf{x}) = a^{2}(t) \hat{\gamma}(\mathbf{x}),
\end{equation}
for spatially homogeneous and isotropic spatial metric tensor $\gamma(t,\mathbf{x})$ and identically vanishing shift vector $ \mathbf{N}(t,\mathbf{x}) $. Here, $ \hat{\gamma}(\mathbf{x}) $ is the metric tensor on the round $2$-sphere, and $a(t)$ is the scale factor. Under these assumptions the momentum constraint \eqref{eq:momconstraint} is trivially satisfied. The metric equations of motion  \eqref{eq:metriceom} are satisfied if and only if
\begin{equation}\label{eq:lambda1eom}
\frac{1}{a(t)}\frac{\mathrm{d}^{2}a(t)}{\mathrm{d}t^{2}} =  \frac{\alpha}{2 \left(2 \lambda-1 \right)}\frac{1}{a^{4}(t)}- \frac{\Lambda}{2 \lambda-1},
\end{equation}
which implies that
\begin{equation}\label{eq:firstint}
\left\{ \frac{\mathrm{d}}{\mathrm{d}t} \left[a^{2}(t)-\frac{\left(2 \lambda-1 \right) C}{2\Lambda} \right] \right\}^{2}= -\frac{2\alpha}{2 \lambda-1}+\frac{C^{2} \left(2 \lambda-1 \right)}{\Lambda} - \frac{4\Lambda}{2 \lambda-1} \left[ a^{2}(t)-\frac{C \left(2 \lambda-1 \right)}{2 \Lambda} \right]^{2}
\end{equation}
for constant of integration $ C $. Letting
\begin{equation}
u(t)=a^{2}(t)-\frac{C \left(2 \lambda-1 \right)}{2 \Lambda},
\end{equation}
\eqref{eq:firstint} becomes
\begin{equation}\label{eq:u}
\left[\frac{\mathrm{d}u(t)}{\mathrm{d}t}\right]^{2}+ \frac{4 \Lambda}{2\lambda-1} u^{2}(t) = -\frac{2 \alpha}{2\lambda-1}+\frac{C^{2}\left(2 \lambda-1 \right)}{\Lambda},
\end{equation}
which has the solution
\begin{equation}\label{usolution}
u(t) = \sqrt{\frac{C^{2}\left(2 \lambda-1 \right)^{2}}{4 \Lambda^{2}}-\frac{\alpha}{2 \Lambda}} \cos \left( 2 \sqrt{\frac{\Lambda}{2 \lambda-1}} t + \delta \right)
\end{equation}
for a second constant of integration $\delta$. In terms of $a(t)$, the solution \eqref{usolution} is
\begin{equation}\label{eq:solelongated}
a_{+}^{2}(t) = \frac{C \left( 2 \lambda-1 \right)}{2 \Lambda} + \sqrt{\frac{C^{2}\left(2 \lambda-1 \right)^{2}}{4 \Lambda^{2}}-\frac{|\alpha|}{2 \Lambda}} \cos \left( 2 \sqrt{\frac{\Lambda}{2 \lambda-1}} t + \delta \right)
\end{equation}
for positive values of $\alpha$ and
\begin{equation}\label{eq:soldeSitter}
a_{-}^{2}(t) = \frac{C \left( 2 \lambda-1 \right)}{2 \Lambda} + \sqrt{\frac{C^{2}\left(2 \lambda-1 \right)^{2}}{4 \Lambda^{2}}+\frac{|\alpha|}{2 \Lambda}} \cos \left( 2 \sqrt{\frac{\Lambda}{2 \lambda -1}} t + \delta \right)
\end{equation}
for negative values of $\alpha$. We assume that $\Lambda>0$ as in the previous studies of causal dynamical triangulations. Note that the solution \eqref{eq:soldeSitter} has a finite time extent dictated by the zeros of $ a_{-}^{2}(t)$ whereas the solution \eqref{eq:solelongated} has no such restrictions. Also note that for vanishing $\alpha$ the solutions \eqref{eq:solelongated} and \eqref{eq:soldeSitter} both reduce to that of Euclidean de Sitter spacetime.

\section{On the Geometry of Causal Dynamical Triangulations}\label{app2}

For the tetrahedra depicted in figure $1$, the Lorentzian dihedral angles between spacelike and timelike faces are
\begin{subequations}\label{angle}
\begin{eqnarray}
\theta_{L}^{(3,1)}&=&\frac{\pi}{2}+ i \log\left(\frac{1+2\sqrt{3 \eta+1}}{\sqrt{3} \sqrt{4 \eta +1 }} \right)\label{eq:theta31} \\
\theta^{(2,2)}_{L} &=& i \log \left( \frac{4\eta+3-2\sqrt{2} \sqrt{2\eta+1}}{4\eta+1} \right) \\
\theta^{(1,3)}_{L} &=& \frac{\pi}{2}+ i \log\left(\frac{1+2\sqrt{3 \eta+1}}{\sqrt{3} \sqrt{4 \eta +1 }} \right),\label{eq:theta22}
\end{eqnarray}
\end{subequations}
and the Lorentzian $3$-volumes are
\begin{subequations}\label{volume}
\begin{eqnarray}
V^{(3,1)}_{L}&=&\frac{1}{12} \sqrt{3 \eta+1} a^{3} \\ V^{(2,2)}_{L} &=& \frac{1}{6 \sqrt{2}} \sqrt{2 \eta+1} a^{3}\\ V^{(1,3)}_{L} &=& \frac{1}{12} \sqrt{3 \eta+1} a^{3}.
\end{eqnarray}
\end{subequations}

As previously mentioned, Wick rotation consists in analytically continuing $\eta$ in the lower half complex plane \cite{DynTri2001}. If the argument of a square root becomes negative as a result of the Wick rotation, then we replace it by the negative of the argument multiplied by $-i$. The Lorentzian dihedral angles \eqref{angle} are thus continued to their respective Euclidean values
\begin{subequations}
\begin{eqnarray}
\theta^{(3,1)}_{E}&=& \frac{\pi}{2}-\cos^{-1} \left( \frac{2 \sqrt{3 \eta-1}}{\sqrt{3} \sqrt{4 \eta-1 }}\right) \\ \theta^{(2,2)}_{E}&=& \cos^{-1} \left( \frac{4 \eta-3}{4 \eta-1} \right) \\ \theta^{(1,3)}_{E} &=& \frac{\pi}{2}-\cos^{-1} \left( \frac{2 \sqrt{3 \eta-1}}{\sqrt{3} \sqrt{4 \eta-1 }}\right) ,
\end{eqnarray}
\end{subequations}
and the Lorentzian $3$-volumes are thus continued to their respective Euclidean values
\begin{subequations}
\begin{eqnarray}
V^{(3,1)}_{E}&=& - \frac{i}{12} \sqrt{3\eta-1} a^{3} \\ V^{(2,2)}_{E} &=& - \frac{i}{6 \sqrt{2}} \sqrt{2 \eta-1} a^{3} \\ V^{(1,3)}_{E}&=& - \frac{i}{12} \sqrt{3\eta-1} a^{3}.
\end{eqnarray}
\end{subequations}
Note that the magnitude of $ \eta $ must be greater than $ \frac{1}{2} $ to prevent the tetrahedra from becoming degenerate.

\end{document}